\documentclass{aa} 
\usepackage{babel}
\usepackage{graphicx}
\usepackage{txfonts}
\usepackage{lipsum}
\usepackage{lscape}             
\usepackage{placeins}           

\newcommand{\tth}{{ {\scshape The Three Hundred}}}

\newcommand{\SfrEfficiency}{$\alpha_{\mathrm{SF}}$}
\newcommand{\FeedbackReheatingEpsilon}{$\varepsilon_{\mathrm{disc}}$}
\newcommand{\FeedbackEjectionEfficiency}{$\varepsilon_{\mathrm{halo}}$}
\newcommand{\ReIncorporationFactor}{$\kappa_{\mathrm{reinc}}$}
\newcommand{\RadioModeEfficiency}{$\kappa_{\mathrm{R}}$}
\newcommand{\QuasarModeEfficiency}{$\kappa_{\mathrm{Q}}$}
\newcommand{\BlackHoleGrowthRate}{$f_{\mathrm{BH}}$}
\newcommand{\ThreshMajorMerger}{$f_{\mathrm{major}}$}
\newcommand{\ThresholdSatDisruption}{$f_{\mathrm{friction}}$}

\newcommand{\EtaSN}{$\eta_{\mathrm{SN}}$}
\newcommand{\modot}{$\mathrm{M}_\odot$}
\newcommand{\h}{$h^{-1}$}

\newcommand{\Alpha}{$\alpha$}
\newcommand{\Epsilon}{$\varepsilon$}
\newcommand{\EpsilonEjec}{$\varepsilon_{\mathrm{Ejec}}$}
\newcommand{\FracReinc}{$f_{\mathrm{reinc}}$}
\newcommand{\PertDist}{$P_{\mathrm{Dist}}$}
\newcommand{\FracBH}{$f_{\mathrm{BH}}$}
\newcommand{\KAGN}{$\kappa_{\mathrm{AGN}}$}
\newcommand{\AlphaRP}{$\alpha_{\mathrm{RP}}$}

\newcommand{\EtaSNsag}{$\eta_{\mathrm{SN}}$}

\begin{document}

    \title{The Three Hundred Project: A fast semi-analytic model emulator of hydrodynamical galaxy cluster simulations}

    \author{
        Jonathan S. G\'omez\inst{\ref{Madrid}, \ref{CIAFF}}
        \and Tomas Hough\inst{\ref{Madrid}, \ref{CIAFF}, \ref{IAP}}
        \and Alejandro Jim\'enez Mu\~noz\inst{\ref{Madrid}, \ref{CIAFF}}
        \and Gustavo Yepes\inst{\ref{Madrid}, \ref{CIAFF}}
        \and Weiguang Cui\inst{\ref{Madrid}, \ref{CIAFF}, \ref{Edimburgo}}
        \and Sof\'ia A. Cora\inst{\ref{IAP}, \ref{FCAG}}
    }

    \institute{
        Departamento de F\'isica Te\'orica, M\'odulo 8, Facultad de Ciencias, Universidad Aut\'onoma de Madrid, 28049 Madrid, Spain. \label{Madrid}\\
        \email{j.s.gomez.u@gmail.com}
        \and
        Centro de Investigaci\'on Avanzada en F\'isica Fundamental (CIAFF), Facultad de Ciencias, Universidad Aut\'onoma de Madrid, 28049 Madrid, Spain. \label{CIAFF}
        \and
        Instituto de Astrof\'isica de La Plata (CCT La Plata, CONICET, UNLP), Paseo del Bosque s/n, La Plata, Argentina. \label{IAP}
        \and
        Institute for Astronomy, University of Edinburgh, Royal Observatory, Blackford Hill, Edinburgh EH9 3HJ, UK. \label{Edimburgo}
        \and Facultad de Ciencias Astronómicas y Geofísicas, Universidad Nacional de La Plata, Paseo del Bosque s/n, La Plata, Argentina. \label{FCAG}        
        }


    \abstract{
        Next-generation photometric and spectroscopic surveys will detect faint galaxies in massive clusters, advancing our understanding of galaxy formation in dense environments. Comparing these observations with theoretical models requires high-resolution cluster simulations. Hydrodynamical simulations effectively resolve galaxy properties in halos; however, they face challenges in simulating low-mass galaxies within massive clusters due to computational limitations. On the other hand, dark matter-only (DMO) simulations can provide higher resolution but need models to populate subhalos with galaxies. In this work, we introduce a fast and efficient emulator of hydrodynamical simulations of galaxy clusters, based on  the semi-analytic models (SAMs) SAGE and SAG. The calibration of the cluster galaxy properties in the SAMs was guided by the cluster galaxies from the hydrodynamical simulations at intermediate resolution, which represents the highest resolution achievable with current hydrodynamical simulations, ensuring consistency in properties such as stellar masses and luminosities across different redshifts. These SAMs are then applied to DMO simulations from {\sc The Three Hundred} Project at three different resolutions. Our results show that the SAG model, unlike  SAGE, more efficiently emulates  the galaxy properties tested in this study even at the highest resolution. This improvement results from the detailed treatment of orphan galaxies, which are satellite galaxies that contribute significantly to the overall galaxy population. SAG enables the study of dwarf galaxies down to stellar masses of $M_* = 10^7 \, \mathrm{M}_\odot$ at the highest resolution, which is an order of magnitude smaller than the stellar masses of galaxies in the hydrodynamical simulations at the intermediate resolution, corresponding to approximately four magnitudes fainter. This demonstrates that a SAM can be effectively calibrated to provide fast and accurate predictions for specific hydrodynamical simulations, offering a computationally efficient alternative for exploring galaxy populations in dense environments across higher resolutions.
        }

    
    \titlerunning{A fast semi-analytic model emulator of hydrodynamical galaxy cluster simulations}
    \authorrunning{Jonathan S. G\'omez et al.}
    \date{Accepted on March 25, 2025}
    \maketitle
    \nolinenumbers

    \section{Introduction}
        \label{section: introduction}

        In the Lambda cold dark matter ($\Lambda$CDM) model, galaxy formation, and evolution are intrinsically linked to the formation and growth of dark matter (DM) halos. Stars originate in cold baryonic gas clouds that condense as hot gas cools. This cooling process occurs due to shocks generated by the gravitational collapse of DM halos \citep{binney1977, rees&ostriker1977, white&rees1978}.

        The formation and evolution of DM halos in $\Lambda$CDM is well understood due to the simplicity of the physics -- to a decent approximation we can assume that DM interacts exclusively through gravity -- which is easily addressed using simulations. However, the evolution of the baryonic component is more unclear and requires choices to be made regarding the subgrid physics (see the review by \citealt{somerville&dave2015} and \citealt{vogelsberger2020}). One of the leading alternatives for modelling the formation and evolution of galaxies in $\Lambda$CDM is semi-analytical modelling (SAM; see e.g. \citealt{cole1991}, \citealt{lacey&silk991} and \citealt{white&frenk1991} for the first examples of such models). This approach uses the evolution of DM halos as obtained from Monte-Carlo prescriptions \citep{kauffmann&white1993, kauffmann1993, lacey&cole1993, cole1994} or N-body simulations \citep{roukema1993, roukema&yoshii1993, roukema1997, kauffmann1999, okamoto&nagashima2001, somerville2008, benson2012} and couples this to simplified physical models of the baryonic physics governing galaxy formation (for reviews, see \citealt{baugh2006} and \citealt{benson2010}).

        From an observational perspective, the next generation of deep surveys—such as {\sc Euclid} \citep{laureijs2011:euclid}, {\sc 4most} \citep{jong2019:4most}, which includes the {\sc Chances} project \citep{sifon2024}, and {\sc Weave} \citep{jin2023:weave}—will detect cluster galaxies down to very faint magnitudes. For instance, {\sc Euclid} will reach limits as low as $m_H \sim 24$ in the $H$-band \citep{jimenez2024}, while {\sc Chances} achieves a photometric depth of $r_\mathrm{AB} \sim 21$ with S-PLUS and selects galaxies spectroscopically up to $r_\mathrm{AB} < 20.5$. These observations will help to understand the physics of galaxy formation and evolution in these high-density environments and will also allow better mass estimates of the total mass of clusters from their galaxy content, which is key to constrain the cosmological parameters of the Universe. At the same time, new methods will also be required to perform cosmological simulations of cluster-size objects with better mass resolutions to be able to contrast the predictions from the theoretical models with those observed by the new surveys.

        The full-physics hydrodynamical simulations of massive clusters generated within {\sc The Three Hundred} \citep{cui2022:gizmo} offer a perfect laboratory for a comparison with current surveys. These hydrodynamical simulations were calibrated using the stellar mass function of satellite galaxies (SSMF) from the most massive clusters to match the observed SSMF of galaxy clusters from \citet{yang2018}, originally derived from the dataset of \citet{yang2012} and based on observations of nearby galaxy clusters in the Sloan Digital Sky Survey (SDSS; \citealt{york2000}). While this calibration improves the agreement between the simulated and observed galaxy populations, they have a mass resolution not enough to resolve DM subhalos below $10^{11}$ \h \modot. Their hosted galaxies have magnitudes that are brighter than those coming from upcoming (e.g. {\sc Euclid}) surveys. For this reason, researchers are developing a new generation of high-resolution hydrodynamical simulations. These high-resolution simulations follow the same calibration methodology as their lower-resolution counterparts in \citet{cui2022:gizmo}, ensuring a consistent framework for comparisons with observational surveys. However, due to their high computational cost, only a limited number of cluster regions have been completed so far. This new set of simulations is not sufficient for statistical studies, but useful for obtaining preliminary results.

        Alternative methods are required to emulate hydrodynamical galaxy cluster simulations and achieve higher-resolution results without the high computational cost of full-physics simulations. One such alternative is to use the semi-analytic models (SAMs) of galaxy formation and evolution on DMO simulations. For this purpose, DMO simulations of {\sc The Three Hundred} dataset at high resolution have already been completed, requiring significantly fewer computational resources. By emulating the observational properties of galaxies within DM halos, we aim to replicate the results from the full-physics hydrodynamical simulations. To achieve this, we employed the SAGE \citep[Semi-Analytic Galaxy Evolution;][]{croton2016:sage} and SAG \citep[Semi-Analytic Galaxies;][]{cora2018} semi-analytic models, calibrating their parameters against synthetic relations extracted from the available hydrodynamical simulations at different redshifts.

        New generations of large-scale cosmological hydrodynamical simulations that contain lots of massive galaxy clusters, such as the FLAMINGO simulation \citep{schaye2023} and the TNG-Cluster simulation \citep{nelson2024}, have been carried out. Additionally, the 324 cluster regions of the hydrodynamical simulations at high resolution from the {\sc The Three Hundred} Project will soon be completed. These simulations operate at three distinct mass resolutions, here referred to as 3K, 7K, and 15K, where the numbers indicate the equivalent of the number of particles per dimension in the original $(1 h^{-1} \rm{Gpc})^3 $ computational volume from which these cluster regions were extracted (see Table \ref{table: the300 dataset}). The 3K resolution corresponds to simulations with lower computational demand but reduced detail in galaxy properties, while the 15K resolution is available only in the DMO simulations at the time of writing this paper, providing the highest level of detail at the cost of requiring substantially more computational resources. The 7K resolution, used for calibration in this work, strikes a balance between computational feasibility and the ability to capture detailed galaxy properties. Therefore, we tested our SAM emulator on DMO simulations with higher mass resolution. Considering that DMO simulations take substantially less time to run than hydrodynamic simulations, this emulator allows us to always be one step ahead in resolution with respect to the full physics simulations.

        The next generation of surveys, such as {\sc Euclid}, {\sc 4most}, and {\sc Weave}, will provide a detailed probe of the galaxy populations in dense environments, particularly through the measurement of the luminosity function in different bands. These surveys will be crucial for addressing several open questions about the properties of low-mass galaxies, such as their star formation rates, morphological evolution, and their role in the assembly of galaxy clusters. High-density environments, such as galaxy clusters, have long been known to influence galaxy properties, but the exact mechanisms are still not fully understood. Models like SAGE and SAG can provide important predictions regarding how these environments impact the evolution of low-mass galaxies, especially in terms of stellar populations, gas content, and feedback processes. With improved resolution from DMO simulations, our ability to predict these properties with accuracy is enhanced, making the connection between theoretical models and future observations a key step in refining our understanding of galaxy formation in the most extreme environments.

        In this work, we present the SAMs (SAGE and SAG) as efficient emulators of hydrodynamical simulations of galaxy clusters. Unlike most previous calibrations, which are typically performed at a single redshift (usually $z=0$), our approach ensures consistency in galaxy properties across multiple redshifts, reflecting a robust treatment of star formation over cosmic time. This methodology not only saves computational time while predicting the results of hydrodynamic runs but also allows us to apply the SAM emulator to a new generation of {\sc The Three Hundred} DMO simulations at 15K resolution, demonstrating consistency across different simulation resolutions.

        The outline of this paper is as follows. In Section \ref{section: data}, we introduce the details of the {\sc The Three Hundred} zoomed simulations available, which are five variants depending on physics and resolutions: full-physics hydrodynamical and DM N-body simulations at 3 resolutions, used to construct the merger trees that are fed into SAGE and SAG semi-analytical model to construct the galaxy population used in this work. In Section \ref{section: sams}, we briefly describe both models and provide a short explanation of each physical parameter used in the calibration of the models. In Section \ref{section: calibration} we show the calibration procedure that was performed in optimizing the free parameters of SAGE/SAG. We show the results of the SAM-emulator compared to the hydrodynamical simulations in Section \ref{section: results}. In Section \ref{section: conclusions}, we summarise and present our conclusions.

    \section{The Three Hundred dataset}
        \label{section: data}

        \begin{table*}
            \normalsize	
            \centering
            \caption{Versions of the {\sc The Three Hundred} simulations used in this work}
            \label{table: the300 dataset}
            \begin{tabular}{ccccccc}
                \hline
                \hline
                          &                     & DM                & DM                        &                    &           &                        \\
                Name      & N particles         & particle mass     & halo resolution           & Gas mass           & N regions & Reference              \\
                          &                     & [\h \modot]       & 100 particles [\h \modot] & [\h \modot]        &           &                        \\
                \hline
                \hline
                3K-DMO    & $3840^3$            & $1.5 \times 10^9$ & $\sim 10^{11}$            & -                  & 324       & \citet{cui2018:the300} \\
                3K-GIZMO  & $2\times 3840^3$    & $1.5 \times 10^9$ & $\sim 10^{11}$            & $2.36 \times 10^8$ & 324       & \citet{cui2022:gizmo}  \\ \hline
                7K-DMO    & $7680^3$            & $1.8 \times 10^8$ & $\sim 10^{10}$            & -                  & 324       & This work              \\
                7K-GIZMO  & $2\times 7680^3$    & $1.8 \times 10^8$ & $\sim 10^{10}$            & $3.0 \times 10^7$  & 150       & In preparation         \\ \hline
                15K-DMO   & $15360^3$           & $2.3 \times 10^7$ & $ \sim10^{9}$             & -                  & 11        & This work              \\ 
                \hline
            \end{tabular}
            \tablefoot{Name: Identifier for each simulation setup. N particles: Total number of particles per component (DM or gas). DM particle mass: Mass of a single DM particle. DM halo resolution: Approximate halo mass corresponding to 100 DM particles. Gas mass: Initial mass of gas particles in hydrodynamical runs. N regions: Number of simulated regions available in this work. Reference: Source of the simulation data. \citet{cui2018:the300} corresponds to the original 3K-DMO simulations; \citet{cui2022:gizmo} to the 3K-GIZMO hydrodynamical runs. The higher-resolution DMO simulations (7K and 15K) were run as part of this work. The 7K-GIZMO simulations are currently under preparation, and their results are presented here for the first time.}
        \end{table*}

        \subsection{The Three Hundred simulations}
            Our dataset is derived from the \tth\footnote{\url{https://www.the300-project.org.}} project \citep{cui2018:the300}, a collection of zoom-in simulations of spherical regions centred on the 324 most massive cluster-sized halos from the DMO MultiDark Planck simulation \citep[MDPL2;][]{klypin2016}. The MDPL2 simulation features a 1 $h^{-1}$Gpc cube containing $3840^3$ dark matter (DM) particles, each with a mass of $1.5 \times 10^9 \, h^{-1}\mathrm{M}_\odot$, and adopts cosmological parameters from \textit{Planck} \citep{planckcollaboration2016}.

            Initial conditions were generated at $z=120$ by identifying the Lagrangian regions of all particles within spherical regions of comoving radius $15 h^{-1}\mathrm{Mpc}$, centred around each of the 324 clusters in MDPL2 at z=0. Within these Lagrangian regions, high-resolution DM and gas particles were populated, while progressively more massive DM particles filled the surrounding areas outside the zoomed region to describe the global gravitational field.
            
            The \tth~Collaboration re-simulated these zoomed regions using five different physics variants (see Section \ref{subsection: avalibale data}), including both hydrodynamical simulations based on the baryonic physics model {\sc Gizmo-Simba} \citep{cui2022:gizmo} and their corresponding DMO simulations. These variants ensure consistent initial conditions and cluster environments across varying resolutions. This 7K-GIZMO run of \tth~ clusters was performed with the new SIMBA-C model \citep{Hough2023}, which is an updated version of SIMBA \citep{Dave2019}.

        \subsection{Available Data}
            \label{subsection: avalibale data}

            In this work, we used five variants of \tth~simulations, summarised in Table~\ref{table: the300 dataset}. These include the 3K-DMO, 3K-GIZMO, 7K-DMO, 7K-GIZMO, and 15K-DMO configurations, which span different combinations of physics and mass resolutions. Readers can refer to the table for detailed specifications of each variant. Once completed, all these simulations were processed with the Amiga Halo Finder \citep[AHF\footnote{http://popia.ft.uam.es/AHF}, ][]{knollmann&knebe2009} to identify halos  within the selected regions. AHF identifies halos using a spherical overdensity algorithm, and each halo can have smaller halos gravitationally bound to it, which are usually referred to as subhalos. Therefore, AHF provides halo catalogues for DMO simulations, and halos, star, and gas particles altogether for hydrodynamic simulations. As galaxies are only naturally present in hydrodynamic runs, we need SAMs to populate DM halos with galaxies. For both DMO and hydrodynamical simulations, different properties are calculated for each DM halo, such as its radius $R_{200c}$, mass $M_{200c}$, and density profile. Here, $R_{200c}$ is the radius within which the mean density of the halo is 200 times the critical density of the Universe, and $M_{200c}$ is the total mass enclosed within this radius. 
            For SAGE, (sub-)halos are connected across different snapshots via the tree builder {\sc MergerTree}, which is bundled with the halo finder AHF. However, for the SAG model, the halo catalogues derived from AHF are processed using {\sc Consistent Trees}, which reconstructs the merger tree by ensuring continuity across snapshots. The final merger trees remain statistically consistent with those generated by {\sc MergerTree}. 
            We combine the halo catalogues and merger trees from the DMO simulations with SAGE and SAG semi-analytic models of galaxy formation and evolution (described in Section \ref{section: sams} and Appendix), in order to obtain catalogues with galaxy properties. 
            For both hydrodynamical simulations and SAMs, different properties are calculated for each galaxy, such as its stellar mass, gas mass, and luminosities for several spectral bands covering from far-UV to radio. The galaxy luminosities are computed using the STARDUST stellar population synthesis model STARDUST \citep{devriendt999:stardust}. The spectral energy distribution (SED) of each galaxy is convolved with the bandpass of each photometric filter to compute the corresponding galaxy luminosity.

            \begin{figure*}
                \centering
                \includegraphics[scale=0.4]{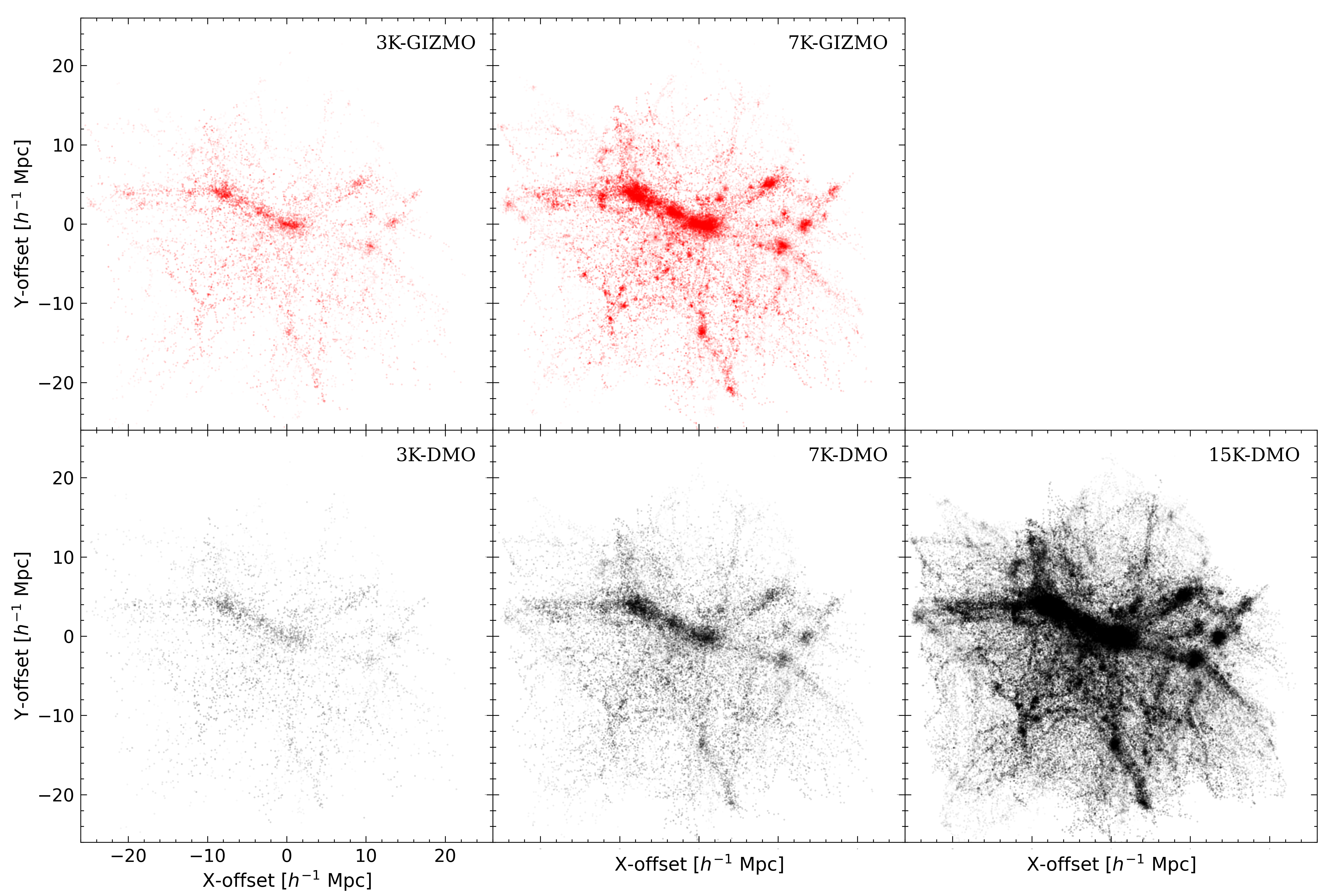}
                \caption{X-Y projection of the entire region 310 in five different flavours depending mainly on the resolution and on the physics used to simulate the region. In the upper panels we show the region in the full-physics hydrodynamic simulations version and in the bottom panels we show the DMO simulations version. The red points represent the dark matter density distribution in the GIZMO simulations. Resolution increases from left to right panels: 3K, 7K, 15K (only DMO) on the left, centre, and right panels respectively.}
                \label{fig: positions}
            \end{figure*}

    \section{Semi-Analytic Model of Galaxy formation and Evolution: SAGE and SAG}
        \label{section: sams}
        
        In this work, we used SAGE and SAG semi-analytic models of galaxy formation and evolution. We describe their general characteristics in the following subsections.
        
        \subsection{SAGE}
            \label{subsection: sage}
            
            The SAGE\footnote{https://github.com/darrencroton/sage} semi-analytic model \citep{croton2016:sage} includes several internal parameters that regulate galaxy formation and evolution, using the merger trees of DM halos as input (mentioned in Section \ref{subsection: avalibale data}). All these parameter values can be tuned to calibrate the SAGE galaxy properties with respect to different (observed or simulated) constraints. In this work, we consider as standard values of these parameters those obtained in \citet{knebe2017} which correspond to a calibration of SAGE performed on a full DM MultiDark simulation \citep{klypin2016}. A summary of these standard parameters can be seen in Table \ref{table: sage parameters}. Furthermore, we carry out a calibration of the SAGE model constrained with an observable from the cluster regions of the hydrodynamic version which will be explained in Section \ref{section: calibration}. A brief description of each parameter considered in this new calibration can be seen in Section \ref{appendix - subsection: sage parameters}. For simplicity, the application of SAGE to the DMO simulations at resolutions of 3K, 7K, and 15K will be referred to as 3K-SAGE, 7K-SAGE, and 15K-SAGE, respectively, from here on.

            In SAGE, when a subhalo is no longer identified in the N-body merger trees, leaving its galaxy without a host subhalo, the galaxy is assumed to merge with the central galaxy of its host halo after a characteristic dynamical friction timescale \citep{croton2016:sage}. The survival time of these galaxies is estimated using empirical relations derived from subhalos with similar properties, where similarity is defined in terms of infall mass, orbital parameters, and host halo characteristics. These relations provide an average merger time for galaxies whose subhalos have disappeared, allowing SAGE to model their eventual fate without explicitly tracking them. Consequently, these galaxies are not evolved between the disappearance of their subhalo and the moment they merge with the central galaxy of the host halo. Thus, they are assumed to remain unchanged during this period.

        \subsection{SAG}
            \label{subsection: sag}
 
            The semi-analytic model of galaxy formation and evolution SAG \citep{cora2006, cora2018} includes usual physical processes: radiative cooling of hot gas, quiescent star formation and starbursts triggered by disc instabilities and galaxy mergers, chemical enrichment, feedback from supernova (SN) explosions, growth of supermassive black holes in galaxy centres and the consequent feedback from active galactic nucleus (AGN). The procedure to obtain the galaxy catalogues within {\sc The Three Hundred} for this model is described in detail in \citet{hough2022}. Similarly to the semi-analytic SAGE model, we calibrated the SAG model optimizing some of its parameters compared to the hydrodynamic simulation version (see Table \ref{table: sag parameters} for standard parameter values obtained in \citealt{knebe2017}). The model parameters used for calibration can be seen in Section \ref{appendix - subsection: sag parameters}. We refer the reader to \citet{cora2018} for a detailed and exhaustive description of the model. For simplicity, the application of SAG to the DMO simulations at resolutions of 3K, 7K, and 15K will be referred to as 3K-SAG, 7K-SAG, and 15K-SAG, respectively, from here on.

            In contrast to the SAGE model, SAG includes a detailed treatment of the orbits of orphan galaxies. Orphan galaxies appear when their subhalos can no longer be identified in the N-body merger trees. Their positions and velocities are derived from a detailed treatment of their orbital evolution, taking into account dynamical friction effects and mass loss as a result of tidal stripping. This treatment allows us to apply the position-based merger criterion and obtain an appropriate radial distribution of satellite galaxies. The orbital evolution of these unresolved subhaloes is tracked in a pre-processing step before applying SAG \citep{delfino2021}.

    \section{SAG and SAGE calibration with Particle Swarm Optimisation}
        \label{section: calibration}

        \begin{figure*}
            \includegraphics[scale=0.296]{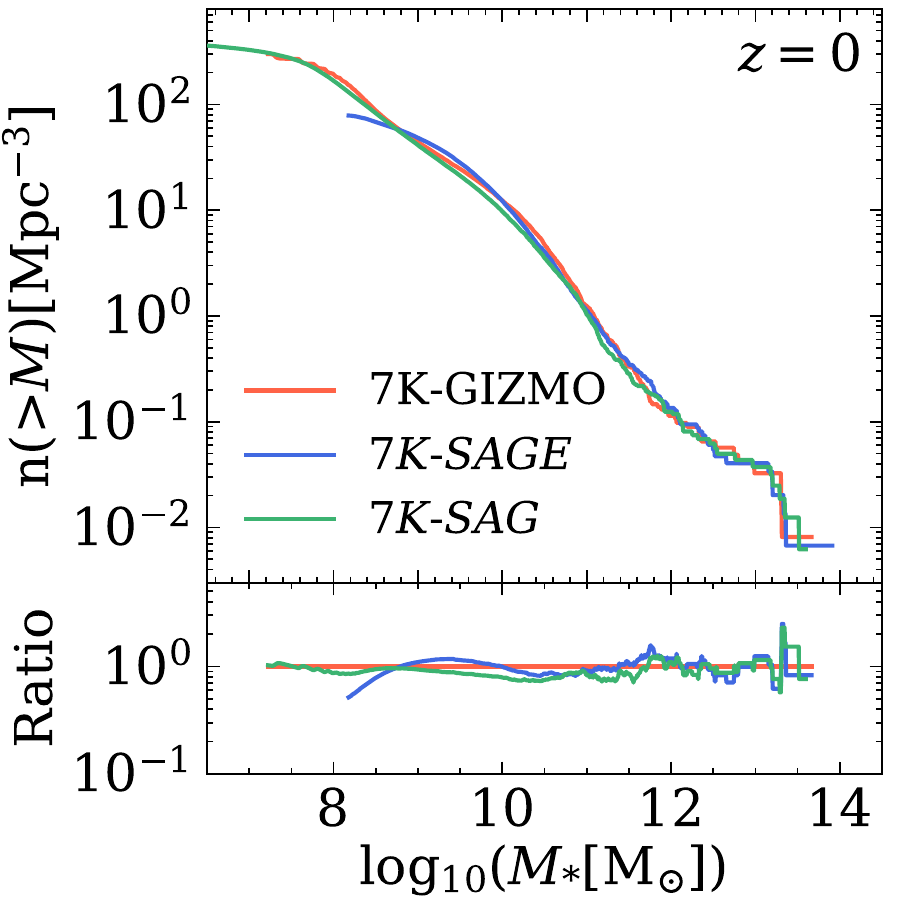}
            \includegraphics[scale=0.296]{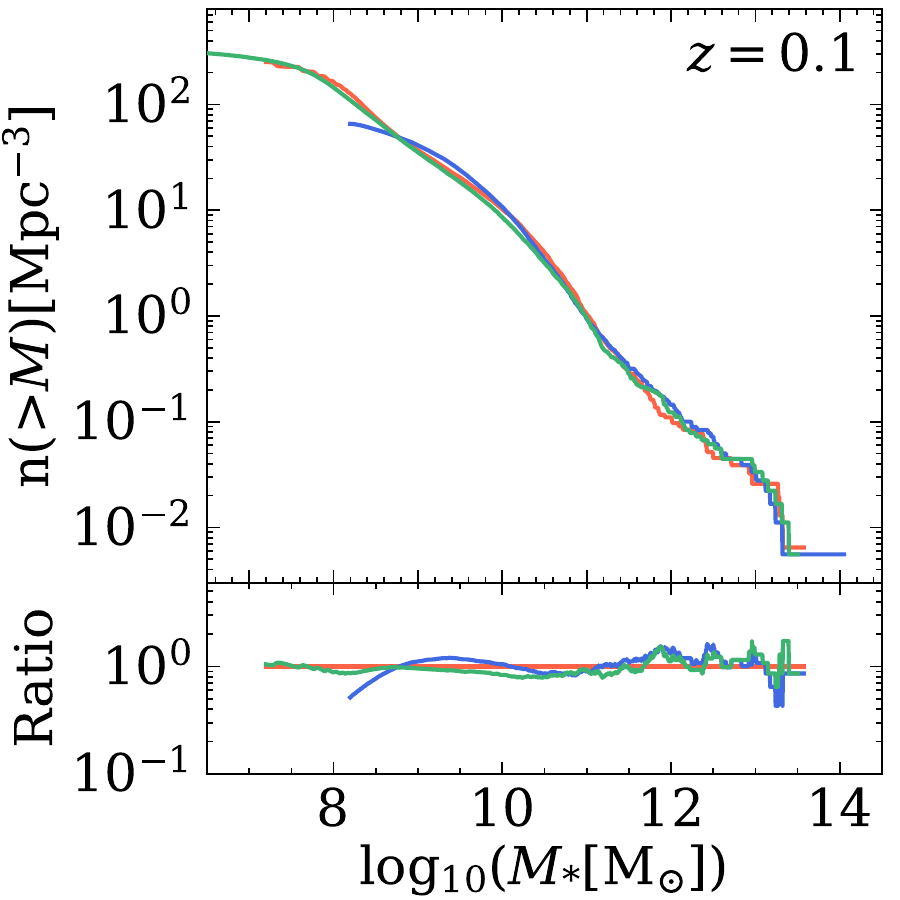}
            \includegraphics[scale=0.296]{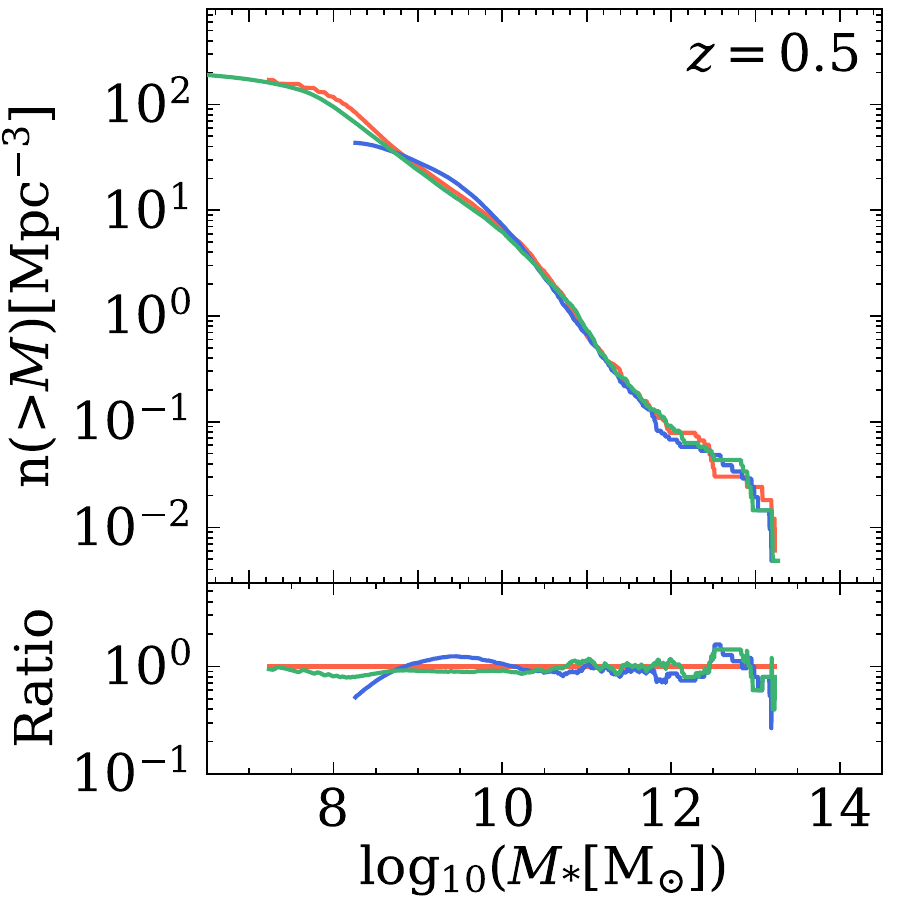}
            \includegraphics[scale=0.296]{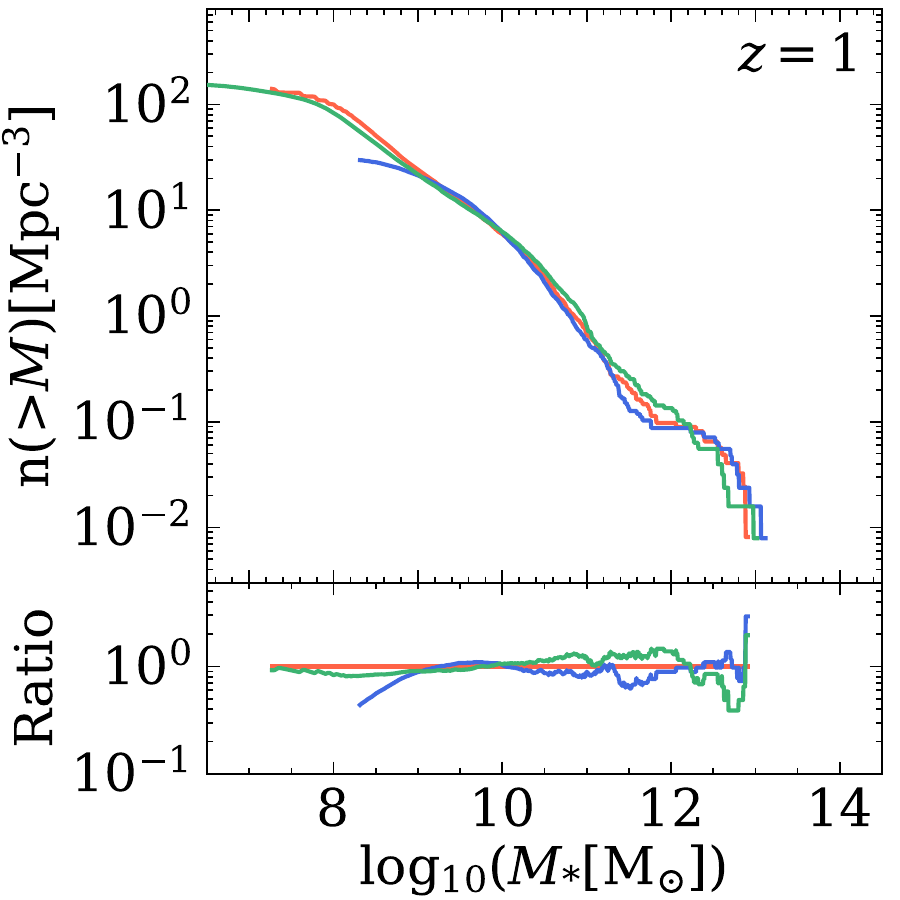}\\
            \includegraphics[scale=0.296]{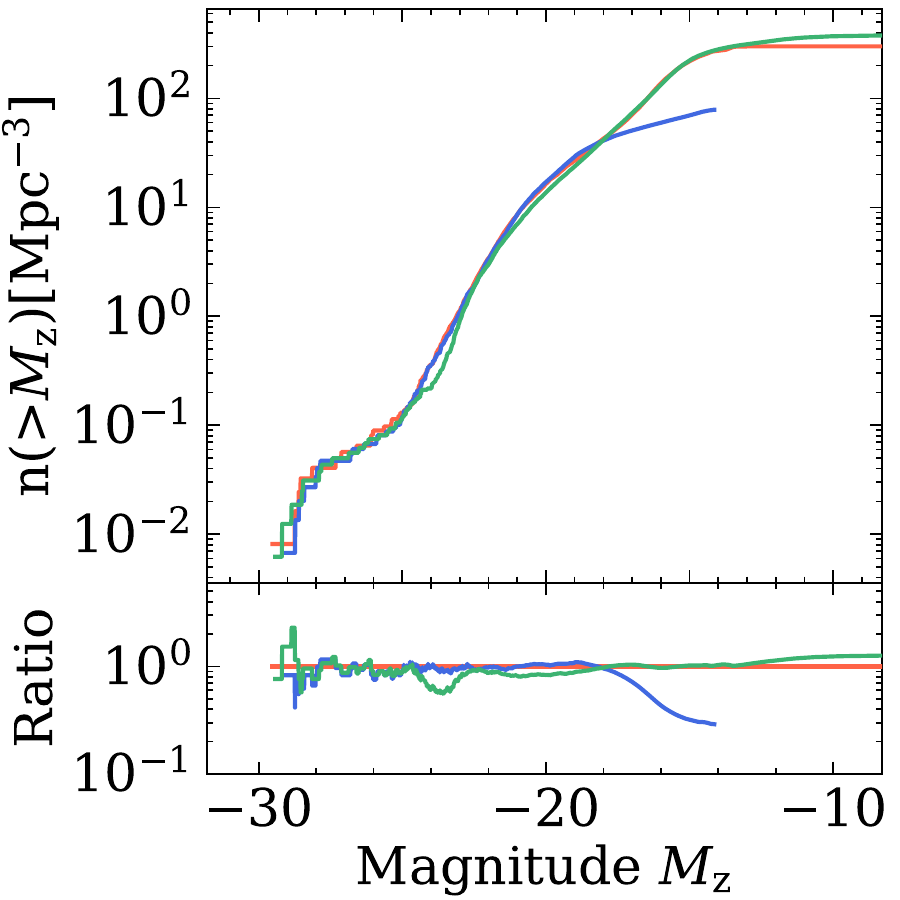}
            \includegraphics[scale=0.296]{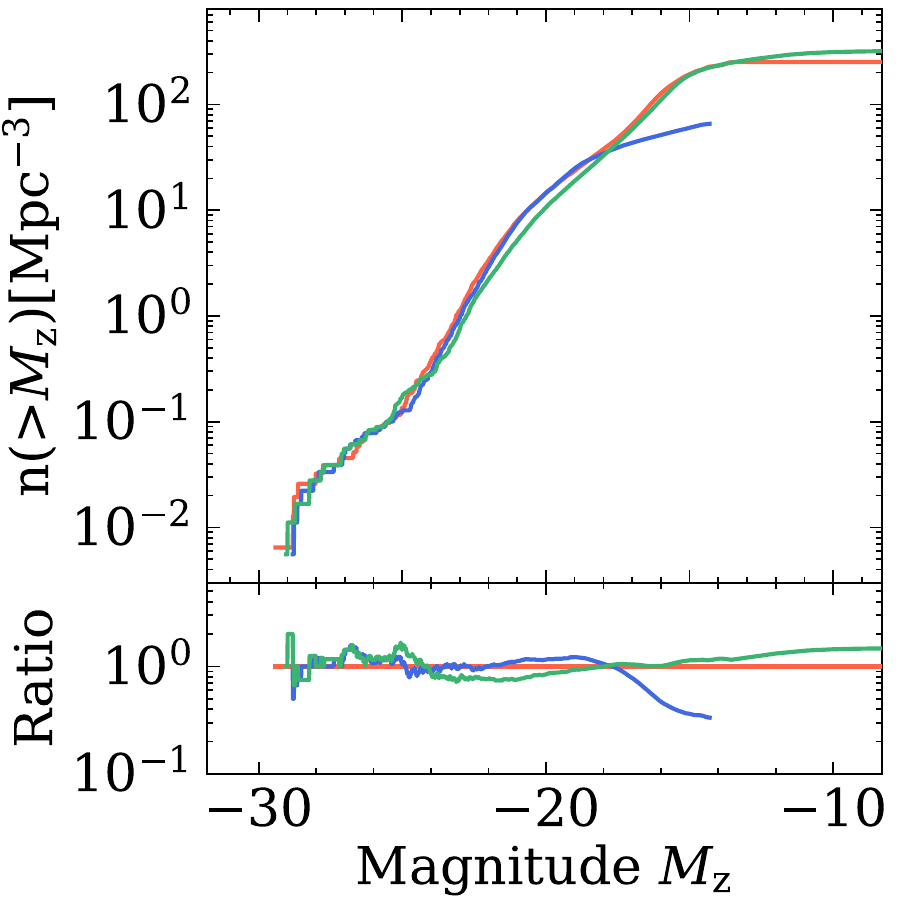}
            \includegraphics[scale=0.296]{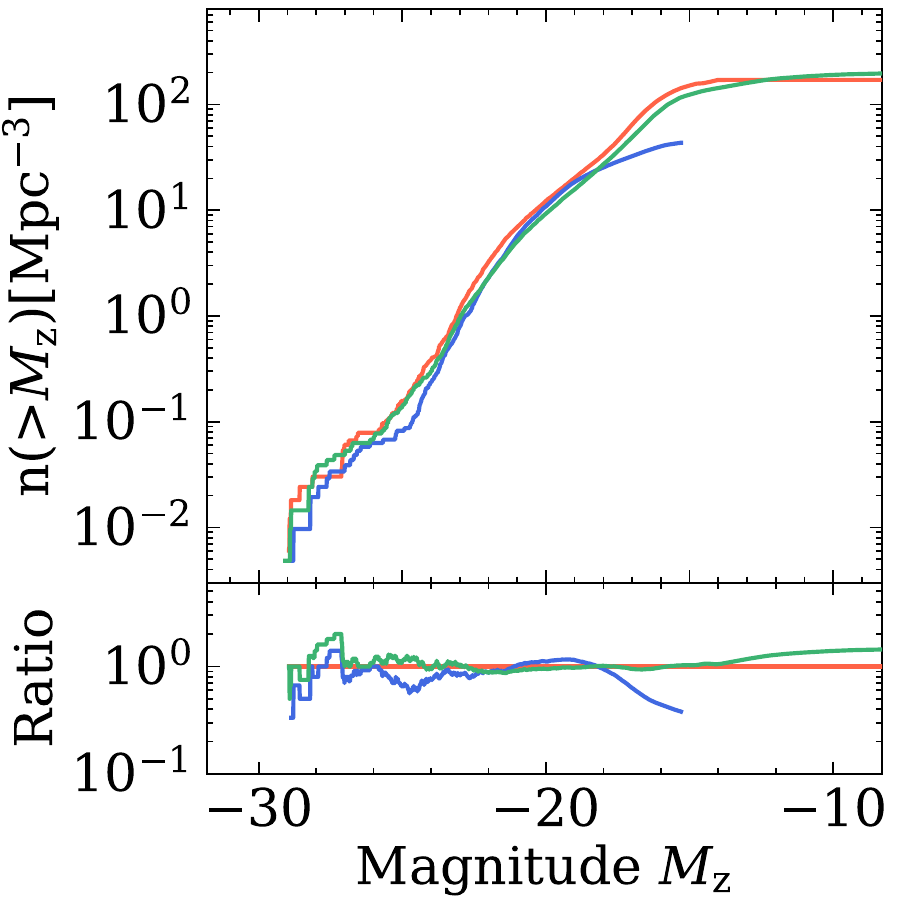}
            \includegraphics[scale=0.296]{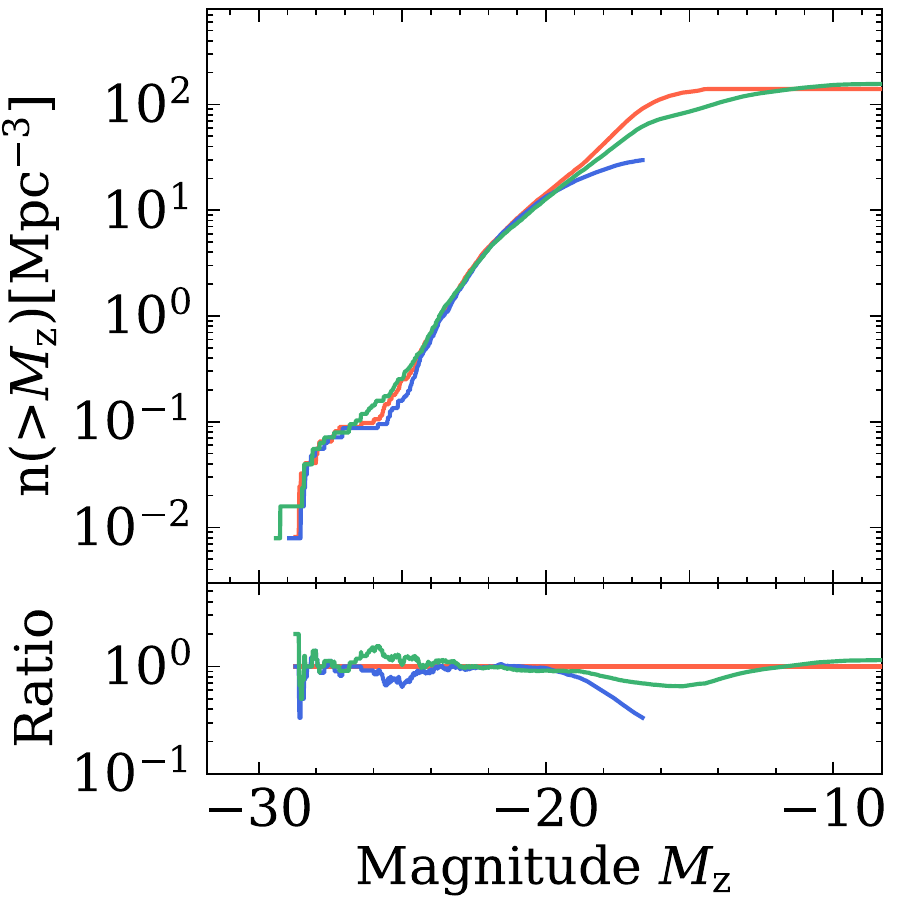}\\
            \includegraphics[scale=0.296]{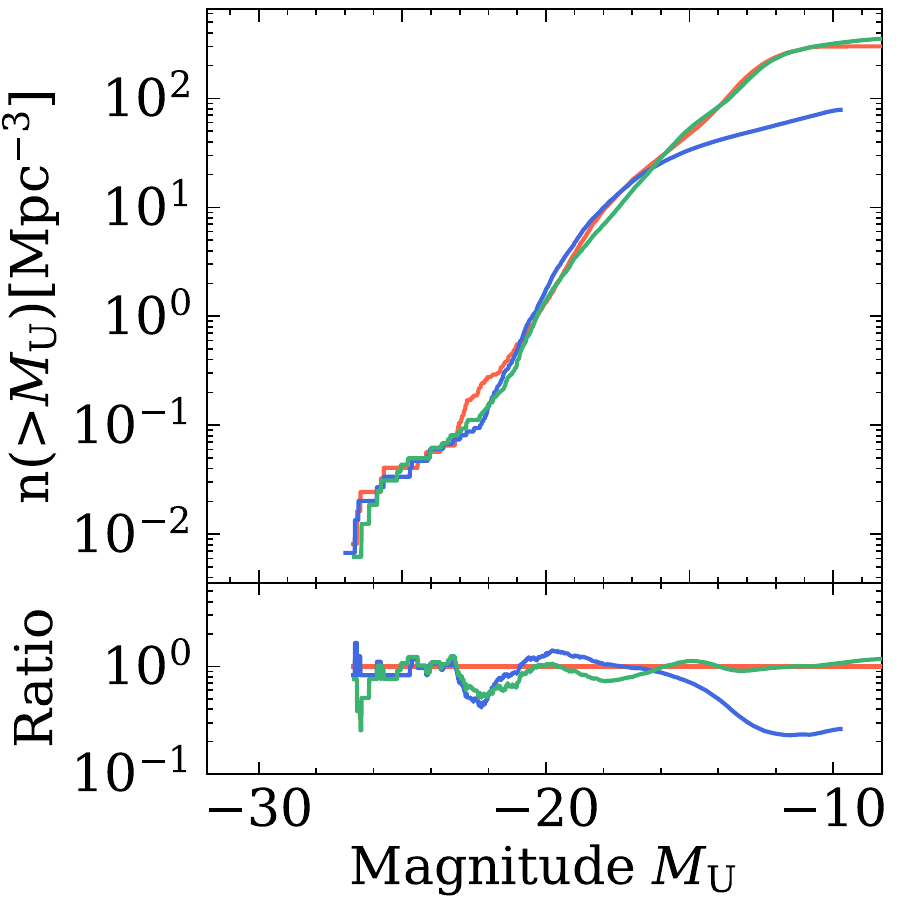}
            \includegraphics[scale=0.296]{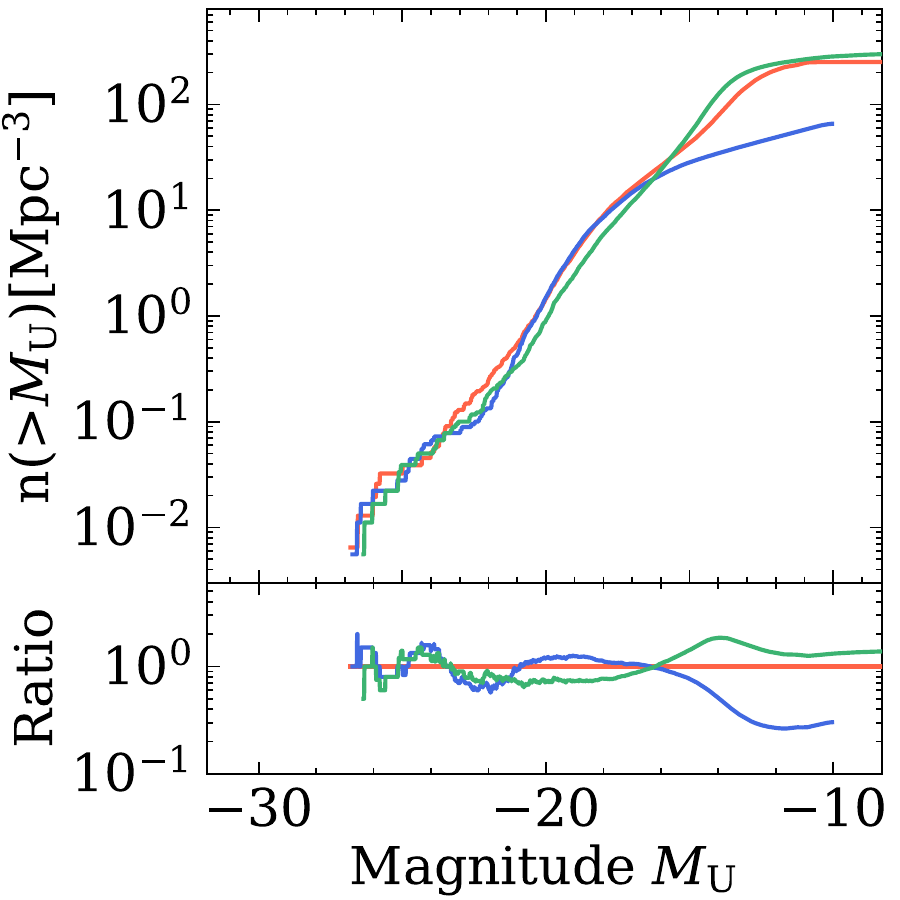}
            \includegraphics[scale=0.296]{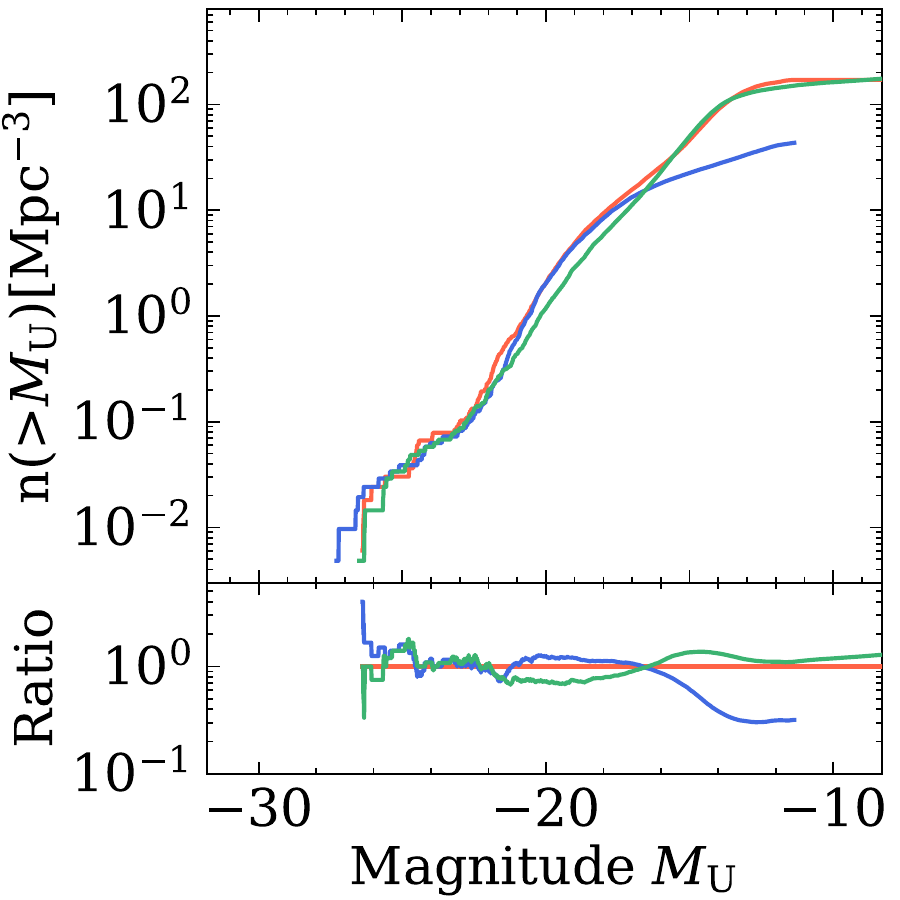}
            \includegraphics[scale=0.296]{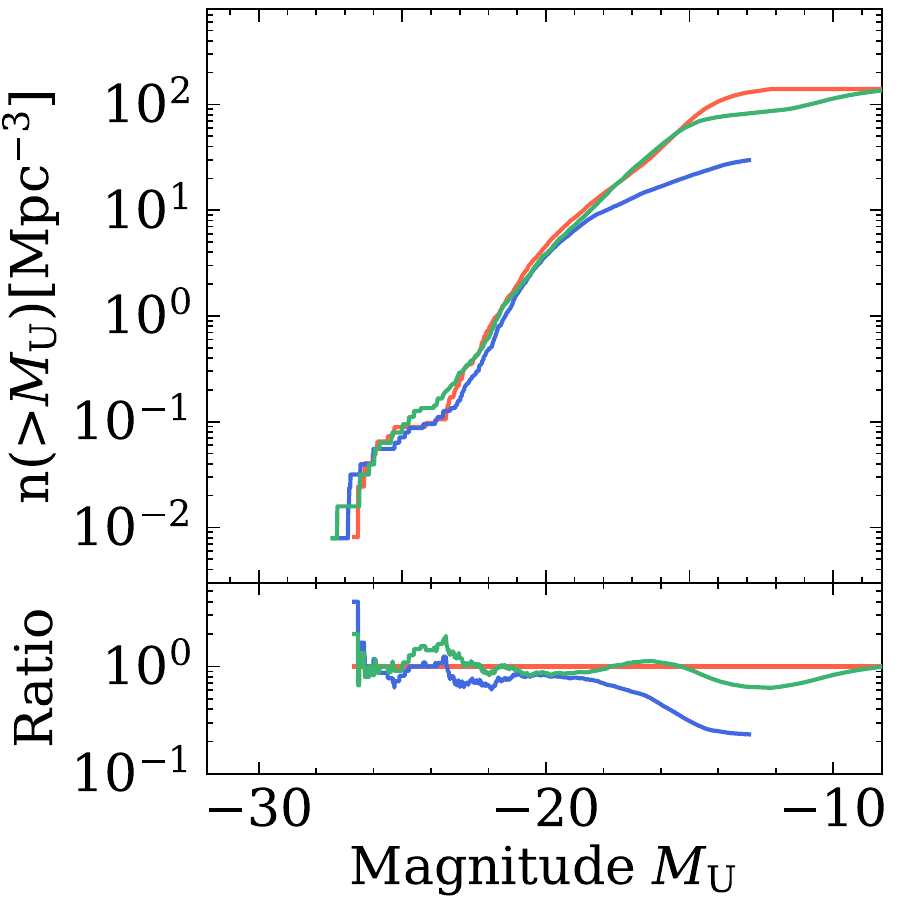}
            \caption{Cumulative stellar mass functions (CSMFs; top panels), cumulative luminosity functions for absolute magnitudes in the $z$-band (CLFs; panels in the second row), and cumulative luminosity functions for absolute magnitudes in the $U$-band (bottom panels) at 4 redshifts: $z=0$, $z=0.1$, $z=0.5$, and $z=1$ (left to right panels). These cumulative functions have been  used to calibrate 7K-SAGE (blue line) and 7K-SAG (green line) with respect to 7K-GIZMO (red line) with PSO using all clusters with $M_{\mathrm{halo}} > 10^{14}$ \modot~ in 5 coincident regions of 7K-DMO and 7K-GIZMO.}
            \label{fig: calibration}
        \end{figure*}

        To calibrate SAGE and SAG, we employed the Particle Swarm Optimisation (PSO) method \citep{eberhart+kennedy1995:pso, kennedy+eberhart1995:pso}, first applied to SAM calibration by \citet{ruiz2015:1°pso+sam}. This method efficiently varies the internal parameters of 7K-SAGE and 7K-SAG (described in Section \ref{appendix - subsection: sage parameters}) to determine the optimal values that minimise the differences between galaxy properties in 7K-SAGE/7K-SAG and those in 7K-GIZMO.

        PSO is an optimisation algorithm inspired by the collective behaviour of social organisms, such as bird flocks or fish schools. Each particle in the swarm represents a possible solution in the parameter space and adjusts its position iteratively based on its own experience and the best-performing solutions of its neighbours. In our case, the multidimensional position of each PSO particle at a given iteration represents the parameter values for a single SAM run, which are updated at each step to minimise the difference between model predictions and hydrodynamical simulation results. This search method is significantly more efficient than traditional Monte Carlo techniques, reducing computational costs by at least a factor of 30 (see \citealt{Kampakoglou2008:sam+mcmc} and \citealt{henriques2009}).
        
        A previous calibration, referred to as `standard' hereafter, presented in \citet{knebe2017}, was performed using the same SAMs applied to the DMO MDPL2 simulation \citep{klypin2016}. Due to the large size of the simulation, the calibration was carried out using merger trees extracted from a smaller box of the MDPL2 simulation, which constitutes a representative sample of the full box, in order to avoid prohibitively long computation times. While these calibrations provided results closely matching observations at low stellar masses, they did not achieve good agreement at high stellar masses, where dense regions dominate. This may be due to the fact that these SAM calibrations compared the results of this smaller box with observations that included galaxies not only from clusters but also from filaments and the field. Thus, the parameters of the SAMs obtained from the standard calibration might not be applicable to the dense DMO regions of the {\sc The Three Hundred} Project. For these reasons, in this work, we performed a new calibration of the SAGE and SAG models, using galaxy clusters at different redshifts, comparing DMO simulations with full-physics hydrodynamical simulations. This makes our calibration the first to be performed on galaxy clusters with consistent temporal evolution.

        \subsection{Calibration method}
            \label{subsection: calibration method}

            \begin{table*}
                \footnotesize
                \centering.
                \caption{Standard and optimized parameters of the SAGE model}
                \begin{tabular}{ccccc}
                    \hline
                    \hline
                    Parameter                   & Description                                                            & Standard value    & Optimized value \\ \hline
                    \hline
                    \SfrEfficiency              & Star formation efficiency                                              & 0.05              & 0.76795         \\
                    \FeedbackReheatingEpsilon   & Mass-loading factor due to supernovae                                  & 3.0               & 2.62696         \\
                    \FeedbackEjectionEfficiency & Efficiency of supernovae to unbind gas from the hot halo               & 0.3               & 0.57674         \\
                    \ReIncorporationFactor      & Sets velocity scale for gas reincorporation                            & 0.15              & 0.12156         \\
                    \RadioModeEfficiency        & Radio mode feedback efficiency                                         & 0.08              & 0.13682         \\
                    \QuasarModeEfficiency       & Quasar mode feedback efficiency                                        & 0.005             & 0.00924         \\
                    \BlackHoleGrowthRate        & Rate of black hole growth during quasar mode                           & 0.015             & 0.00424         \\
                    \ThreshMajorMerger          & Threshold mass ratio for merger to be major                            & 0.3               & 0.30224         \\
                    \ThresholdSatDisruption     & Threshold subhalo-to-baryonic mass for satellite disruption or merging & 1.0               & 0.82527         \\
                    \hline
                \end{tabular}
                \tablefoot{The standard parameter values were obtained from the calibration of SAGE on a smaller, representative volume of the MDPL2 simulation \citep{knebe2017}. The optimized values correspond to our calibration on The Three Hundred 7K-DMO simulations using particle swarm optimisation.}
            \label{table: sage parameters}
            \end{table*}

            \begin{table*}
                \footnotesize
                \centering
                \caption{Standard and optimized parameters of the SAG model}
                \begin{tabular}{ccccc}
                    \hline
                    \hline
                    Parameter                   & Description                                                                        & Standard value          & Optimized value     \\ \hline
                    \hline
                    \Alpha                      & Star formation efficiency                                                          & 0.04026                 & 0.08395             \\
                    \Epsilon                    & SNe feedback efficiency in bulge and disc                                          & 0.32991                 & 0.01257             \\
                    \EpsilonEjec                & The efficiency of ejection of gas from the hot phase                               & 0.02238                 & 0.03722             \\
                    \FracReinc                  & Fraction of ejected reheated cold gas that is reincorporated into the hot halo gas & 0.05554                 & 0.03023             \\
                    \PertDist                   & Factor involved in the distance scale of perturbation to trigger disc instability  & 14.5571                 & 34.2456             \\
                    \FracBH                     & Fraction of cold gas accreted onto the central SMBH                                & 0.06058                 & 0.14338             \\
                    \KAGN                       & Efficiency of cold gas accretion onto the SMBH during gas cooling                  & $3.02198\times 10^{-5}$ & $2.8\times 10^{-4}$ \\
                    \AlphaRP                    & Ram pressure stripping efficiency                                                  & 5.0                     & 8.02785             \\
                    \EtaSNsag                   & The number of SNe generated from the stellar population                            & 6.3e-3                  & 0.01813             \\
                    \hline
                \end{tabular}
                \tablefoot{The standard values were obtained from the calibration of SAG on a smaller, representative volume of the MDPL2 simulation \citep{knebe2017}. In this work, we calibrated \texttt{\AlphaRP} and \texttt{\EtaSNsag}, which were not treated as free parameters in the standard calibration, to improve the model's agreement with the target simulation.}
            \label{table: sag parameters}
            \end{table*}

            From the 150 available 7K-GIZMO regions, we select five that represent different environments and also exist in their 7K-DMO version to perform the calibration method. This is the minimum necessary to ensure a successful calibration process with PSO, as determined through testing with fewer regions, where good calibrations were not achieved. This approach helps us make the most of the available data and accelerates the calibration process, as described in \citet{henriques2009, henriques2013}. Once the SAM models are calibrated, we apply them to the full 7K-DMO dataset to generate the galaxy catalogues.

            As we want to reproduce the star formation history of the 7K-GIZMO runs, we take as calibrators the cumulative stellar mass function (CSMF) and cumulative luminosity functions (CLFs) for the absolute magnitude in the $z$-band and $U$-band of galaxies within all clusters more massive than $10^{14}$ \h \modot. We compute these observables from the five regions of 7K-GIZMO simulations and also compute them in their corresponding DMO version 7K-DMO simulations using SAGE and SAG. The galaxy luminosities were calculated using the STARDUST model \citep{devriendt999:stardust} for both 7K-SAM and 7K-GIZMO. We calibrated the free parameters of the SAMs to simultaneously fit these three galaxy relations at four different redshifts: $z=0$, $z=0.1$, $z=0.5$, and $z=1$, in order to ensure a consistent evolution of these properties relative to the hydrodynamical results.

            For each SAM model, a large number of parameters were changed, 14 parameters for SAGE and 13 parameters for SAG. These internal parameters are closely related to the constraints selected for calibration to ensure optimal results. Additionally, the ranges used for each parameter in the calibration were rigorously selected to have parameter values with consistent physics. We show nine free parameters (the most related to the selected hydrodynamic constraints) in summary Tables \ref{table: sage parameters} and \ref{table: sag parameters}, and a brief explanation and the selected ranges of the parameters in the Appendix \ref{appendix - seccion: sam parameters}.

            We calculate the likelihood of the samples by computing their $\chi^2$, which corresponds to the $\chi^2$ obtained in each comparison function between 7K-SAGE/7K-SAG and 7K-GIZMO, 12 individual $\chi^2$ in total. Then, we minimise with PSO the global likelihood $\chi^2_G$ corresponding to the sum of individual $\chi^2$ obtained from each constraint. The algorithm stops when the best-fitting value does not change significantly for at least 100 steps suggested in the original PSO algorithm \citep{eberhart+kennedy1995:pso, kennedy+eberhart1995:pso}.

            In Fig. \ref{fig: calibration}, we show the final calibration of 7K-SAGE and 7K-SAG. For SAGE, we set a minimum stellar mass of $M_* \geq 10^{9}$ \modot~ and a maximum absolute magnitude limit of $M_z \leq -18$ and $M_U \leq -16$ when calibrating with PSO. This choice is due to the lower number of DM halos in the 7K-DMO simulations compared to 7K-GIZMO, as hydrodynamical simulations of galaxy clusters contain more low-mass subhalos, which are better able to survive due to the presence of baryonic matter \citep{dolag2009}. This can be seen in Fig. \ref{fig: chmf, 7k and 15k, z=0} in which we show the cumulative halo mass function (CHMF) of DM clusters more massive than $M_\mathrm{cluster} < 10^{14}$ \h \modot~ for matched regions in GIZMO and DMO simulations at the 3 available resolutions. DMO simulations (black lines) have fewer halos with $M_\mathrm{halo} < 10^{12}$ \modot~ than the GIZMO run (red lines) at the same resolution simulation, although they reach the same lower limit for the halo mass in both. This effect can even be seen graphically in an XY position projection of a region at different resolutions in Fig. \ref{fig: positions} where the DMO simulations underestimate the number density of halos in the centres of clusters relative to hydrodynamical simulations. This effect was found to be stronger in denser regions \citep{haggar2021}. This suggests that a SAM which does not treat orphan galaxies will never be able to generate the same population of galaxies as a hydrodynamic simulation from the DMO simulation version. Thus, since we have the same halo mass limit in DMO and hydrodynamic simulations at the same resolution, in the case of SAGE we have a limit given by the total number of subhalos in a cluster in the DMO simulation version, implying that we force the functions we are calibrating within the mentioned stellar mass and magnitude limits. On the other hand, in the case of 7K-SAG, the limit is imposed by the 7K-GIZMO hydrodynamic simulation, considering a stellar mass cut of $M_* > 10^{8}$ \modot~ and maximum magnitude limit of $M_z < -15$ and $M_U < -13$, reaching one order of magnitude lower in stellar mass and 3 magnitudes fainter than 7K-SAGE.

        \subsection{Calibration results}

            In Fig. \ref{fig: calibration}, we present the constraint functions obtained from 7K-GIZMO and the corresponding results from the 7K-SAGE and 7K-SAG models. These include the cumulative luminosity functions (CLFs) in the z-band and U-band, and the cumulative stellar mass function (CSMF) at the four redshifts used for calibration. Overall, the calibrated functions show good agreement with 7K-GIZMO. However, while 7K-SAGE reproduces the high-mass end of the CSMF more accurately at high redshift, 7K-SAG achieves a better fit at the low-mass end, thanks to the inclusion of orphan galaxies. The final values of the calibrated parameters for SAGE and SAG are presented in Tables \ref{table: sage parameters} and \ref{table: sag parameters}, respectively. For both SAM models, in general, the optimized values are higher than the standard values. For instance, considering the extreme conditions of the dense regions studied, it is not unusual to find a higher star formation efficiency (\SfrEfficiency~parameter for SAGE, and \Alpha~parameter for SAG) as suggested in \citet{ghodsi2024}. 
            
            For SAG, we further illustrate the likelihood distributions of the calibrated parameters in Fig. \ref{fig: sam parameters sag: standard and calibrated}. Notably, the likelihood distributions show well-defined peaks for most parameters, indicating a stable optimisation process. This stability suggests that the calibrated parameters are robust to variations within the defined ranges and that the chosen parameter values maximize the agreement with 7K-GIZMO. While some parameters exhibit sharp likelihood peaks, reflecting their critical roles in regulating key physical processes, others display broader peaks, indicating a wider range of acceptable values that still yield good agreement. This behaviour aligns with the understanding that certain processes, such as disc instability and starbursts, are less sensitive to small parameter variations compared to processes directly affecting mass or energy regulation, such as SNe feedback. The stability of the optimized parameters within these likelihood distributions highlights the reliability of the SAG calibration process. It also ensures that the model remains consistent when applied to other regions or higher-resolution simulations, further validating the robustness of SAG as an emulator for hydrodynamical simulations.

    \section{Results}
        \label{section: results}

        In this Section, we study how the SAGE and SAG re-calibrated models can reproduce and predict the behaviour of the GIZMO simulations at different numerical resolutions: 3K, 7K, and 15K. The SAM calibration was performed by comparing them with GIZMO at the same 7K resolution (see Sect.~\ref{section: calibration}). The SAM galaxy catalogues at 3K and 15K resolutions were constructed using the parameters obtained in the 7K calibration. Thus, we assess whether the results of the SAM calibration, obtained using only five regions, can be extrapolated to the 150 matching regions between 7K-GIZMO and 7K-DMO in Sect.~\ref{subsection: 5 and 150 regions comparison}. Additionally, we evaluate whether the galaxy properties derived from the calibrated SAMs are sensitive to the resolution of the DM halos in Sect.~\ref{subsection: 7K and 15K comparison}. Finally, we predict other properties of the galaxy cluster and compare them with observations of galaxy clusters in Sect.~\ref{subsection: prediction of galaxy properties}.

        \subsection{Impact of the number of simulated regions on models calibration}
            \label{subsection: 5 and 150 regions comparison}

            \begin{figure}
                \includegraphics[scale=0.59]{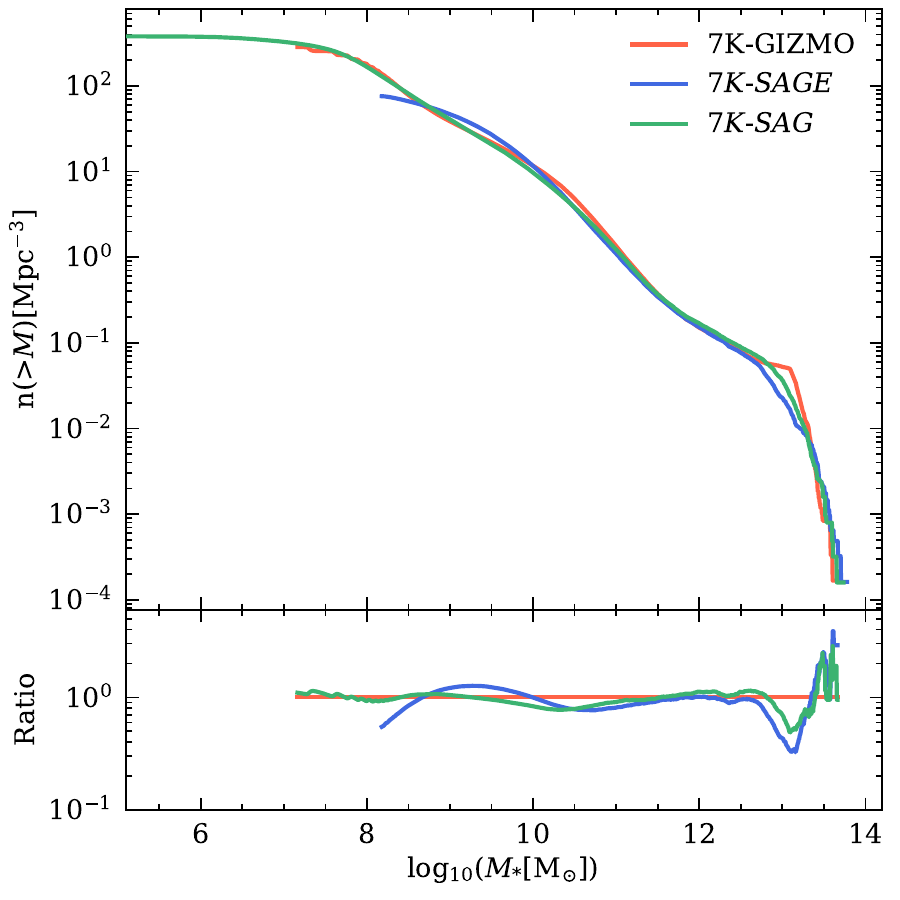}
                \caption{The cumulative halo mass function at $z=0$ obtained by stacking the cumulative functions of all halo clusters with $M_{\mathrm{halo}} > 10^{14}$ \modot~ in the 150 coincident regions between 7K-GIZMO, 7K-SAGE, and 7K-SAG. The bottom panel shows the ratio compared with 7K-GIZMO.}
                \label{fig: csmf, 150regions, z=0}
            \end{figure}

            The first question we aim to address is whether the SAM models, calibrated using only five selected regions, can accurately reproduce the galaxy properties of the 150 overlapping regions between 7K-GIZMO and 7K-DMO, thereby enabling predictions of the statistical behaviour for the full set of 324 DMO regions. For this analysis, we selected the overlapping dark matter clusters between 7K-GIZMO and 7K-SAGE/7K-SAG with \( M_{\mathrm{halo}} > 10^{14} \, \mathrm{M}_\odot \) at \( z=0 \), derived from the 150 available regions. This selection, representing nearly 300 galaxy clusters, is consistently applied across all figures in this subsection.

            The cumulative stellar mass functions (CSMFs) of these clusters are shown in Fig. \ref{fig: csmf, 150regions, z=0}. The 7K-GIZMO model (red line) establishes the minimum stellar mass limit, $M_{\mathrm{*}} = 10^8 \, \mathrm{M}_\odot$, and follows a power-law-like shape in the CSMF. This power-law behaviour extends between $M_{\mathrm{*}} \sim 10^8 \, \mathrm{M}_\odot$ and $M_{\mathrm{*}} \sim 10^{13} \, \mathrm{M}_\odot$, where the slope remains approximately constant. Below $M_{\mathrm{*}} \sim 10^8 \, \mathrm{M}_\odot$, the curves are limited by resolution effects. At the high-mass end, the slope changes due to the transition from central galaxies, which dominate this range and exhibit higher stellar masses, to satellite galaxies, which contribute significantly at lower stellar masses. This transition manifests as a pronounced peak in the CSMF, reflecting the relative contributions of these two populations. In contrast, the SAMs (7K-SAGE and 7K-SAG) exhibit a smoother transition between central and satellite galaxies, lacking the pronounced peak observed in 7K-GIZMO, likely due to their simplified treatment of galaxy populations. The 7K-SAGE model (blue line) shows general agreement with 7K-GIZMO but does not reach the same minimum stellar mass limit for galaxies. This is primarily due to the absence of orphan galaxies in SAGE, which leads to a lower abundance of low-mass galaxies compared to hydrodynamical simulations. Additionally, the lack of baryonic physics in dark matter-only simulations causes an underestimation of low-mass halos \citep{haggar2021}, further limiting the number of small galaxies that SAGE can generate. On the other hand, the 7K-SAG model closely follows the CSMF of 7K-GIZMO, reaching and even extending to lower stellar masses, thanks to its treatment of orphan satellite galaxies. For both 7K-SAMs, slight differences appear at high stellar masses ($M_{\mathrm{*}} > 10^{13} \, \mathrm{M}_\odot$), likely due to statistical fluctuations given the small number of high-mass galaxies present in the 5 calibration regions. While this explanation is a valid option, another source of error could stem from the inherent difficulty of SAMs in accurately calibrating high stellar masses. Previous works, such as \citet{ruiz2015:1°pso+sam, knebe2017, cora2018, cui2018:the300}, demonstrate that SAM calibrations often struggle to reproduce the high-mass end of the SMF in cosmological simulations when compared to hydrodynamical simulations and observations. In this work, we calibrated directly with the dense regions of galaxy clusters, which reduces the discrepancies at high stellar masses, achieving a significantly improved calibration for massive galaxies compared to previous studies. Among the SAM models, 7K-SAG remains the closest to 7K-GIZMO in this stellar mass range.

            \begin{figure}
                \includegraphics[scale=0.59]{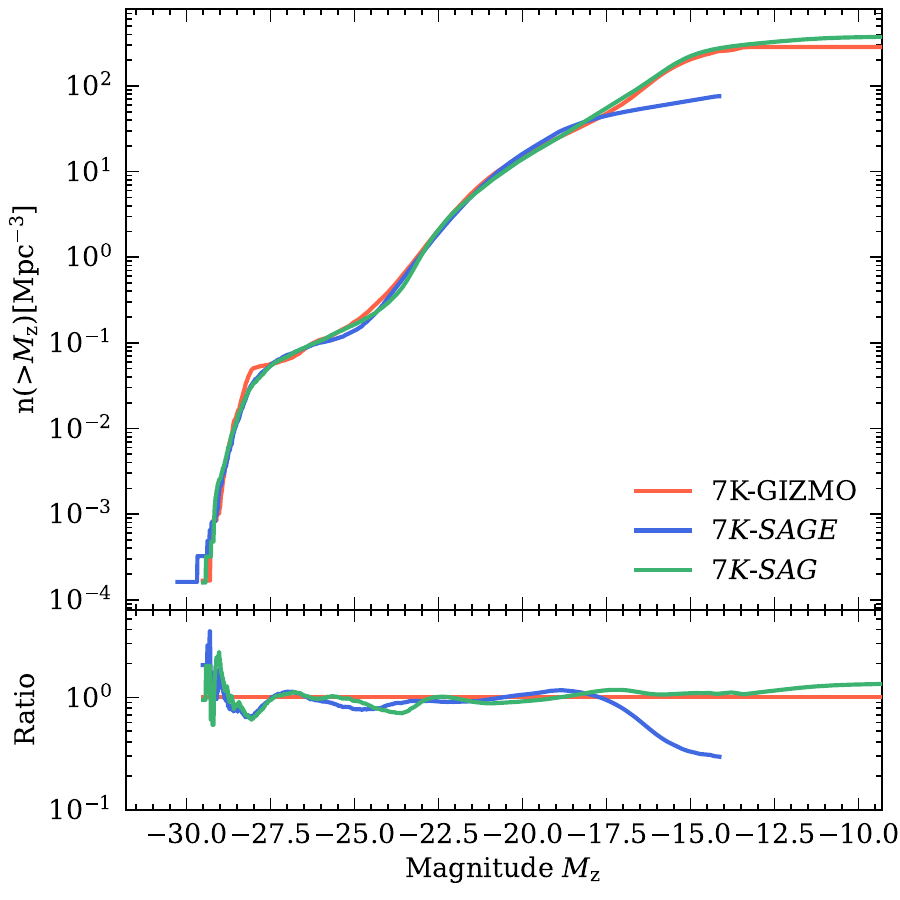}
                \caption{Cumulative luminosity function (CLF) for the absolute magnitude in the $z$-band at $z=0$, derived from all halo clusters with $M_{\mathrm{halo}} > 10^{14} M_\odot$ in the 150 coincident regions between 7K-GIZMO, 7K-SAGE and 7K-SAG, where the $z$-band was one of the bands employed in the calibration of the 7K-SAM. The bottom panel shows the ratio of all the CLFs compared with that of 7K-GIZMO.}
                \label{fig: clf, 150 regions, z=0}
            \end{figure}

            In Fig. \ref{fig: clf, 150 regions, z=0}, we present the cumulative luminosity function (CLF) for the absolute magnitude in the $z$-band at $z=0$, derived from all coincident halo clusters with $M_{\mathrm{halo}} > 10^{14} \, \mathrm{M}_\odot$ at 7K resolution for the SAGE, SAG, and GIZMO simulations. The $z$-band, one of the bands employed in the calibration of the 7K-SAM models, acts as a proxy for stellar mass content, allowing us to draw similar conclusions to those from Fig. \ref{fig: csmf, 150regions, z=0}: both the 7K-SAGE and 7K-SAG models show general agreement with 7K-GIZMO. However, unlike 7K-SAGE, the 7K-SAG model reaches lower luminosities than 7K-GIZMO before the cumulative function flattens. At the high-luminosity range ($M_{\mathrm{z}} \sim -28$) dominated by central galaxies, the ratio of the CLF becomes noisy, reflecting the same behaviour observed at high stellar masses ($M_{\mathrm{*}} > 10^{13} \, \mathrm{M}_\odot$) in the CSMF. This noise is consistent with the low number of regions used in the calibration and the challenge of aligning SAM predictions with hydrodynamical simulations for massive galaxies. Additionally, in this range, 7K-GIZMO shows a pronounced peak caused by the contribution of central galaxies, which is not present in the SAMs, where the transition between central and satellite galaxies is smoother.

            \begin{figure}
                \includegraphics[scale=0.59]{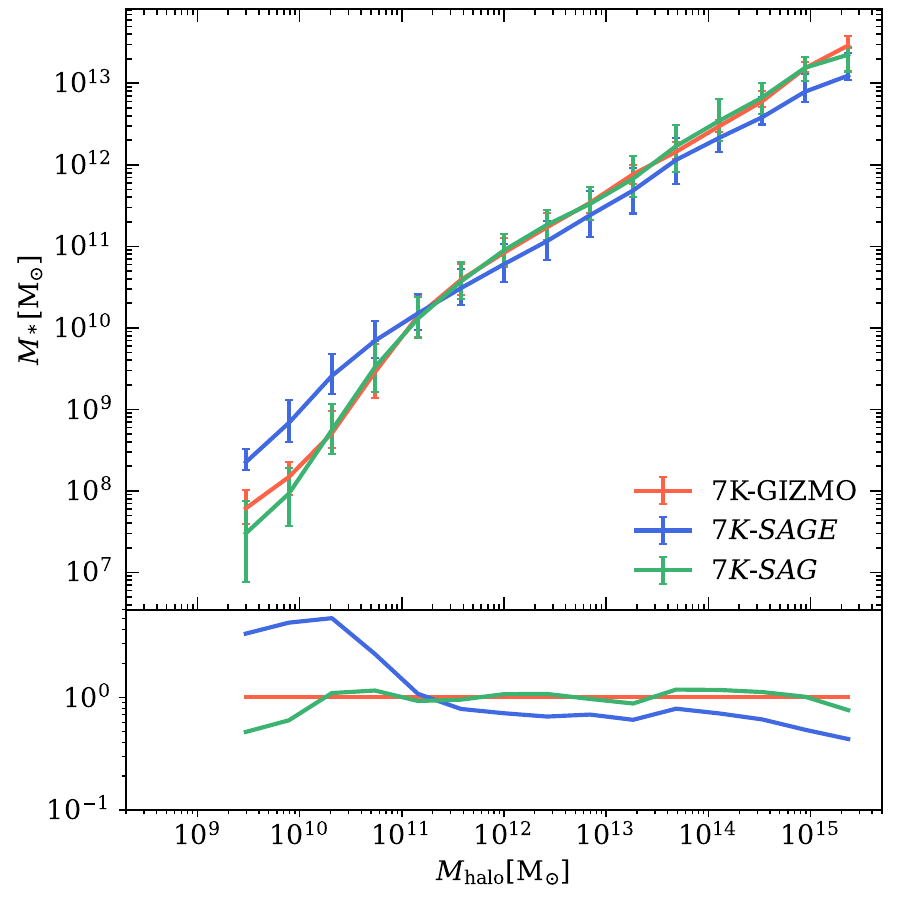}
                \caption{Stellar mass - halo mass ($M_{\mathrm{*}} - M_{\mathrm{halo}}$) relation at $z=0$ for all halo clusters with $M_{\mathrm{halo}} > 10^{14} \, \mathrm{M}_\odot$ in the 150 coincident regions between 7K-GIZMO, 7K-SAGE, and 7K-SAG. Here, $M_{\mathrm{halo}}$ refers to both host halos and subhalos, encompassing central and satellite galaxies. The lines represent the median values, while the error bars denote the 25th and 75th percentiles.}
                \label{fig: smass vs hmass, 150 regions, z=0}
            \end{figure}
            
            While the magnitudes and stellar masses of the SAMs are consistent with GIZMO, the halo mass-stellar mass ($M_{\mathrm{halo}} - M_{\mathrm{*}}$) relation does not necessarily match. In Fig. \ref{fig: smass vs hmass, 150 regions, z=0}, we present the $M_{\mathrm{halo}} - M_{\mathrm{*}}$ relation for all coincident galaxy clusters for the three models. Here, $M_{\mathrm{halo}}$ includes both host halos (central galaxies of clusters) and subhalos (satellite galaxies), allowing for a complete comparison across different models. The 7K-GIZMO model exhibits a clear transition in the slope of the $M_{\mathrm{halo}} - M_{\mathrm{*}}$ relation around $M_{\mathrm{halo}} \sim 10^{12} \, \mathrm{M}_\odot$. This transition reflects the dominance of central galaxies in massive halos and the increasing contribution of satellite galaxies at lower halo masses. The shape of the relation is relatively smooth across the entire range, with stellar masses steadily increasing with halo mass. At lower halo masses ($M_{\mathrm{halo}} < 10^{11} \, \mathrm{M}_\odot$), the 7K-SAGE model significantly overestimates the stellar mass compared to 7K-GIZMO, being approximately three times higher in this range, while 7K-SAG remains more consistent with the hydrodynamical simulations. This is likely due to the detailed treatment of orphan galaxies in SAG, which improves its ability to reproduce the stellar mass content in low-mass halos. At higher halo masses ($M_{\mathrm{halo}} > 10^{12} \, \mathrm{M}_\odot$), 7K-SAG closely follows 7K-GIZMO, whereas 7K-SAGE underestimates the stellar mass content, being approximately $50\%$ lower than 7K-GIZMO, potentially due to limitations in the calibration of massive galaxies. This behaviour is better reproduced by 7K-SAG than by 7K-SAGE. Thus, with SAG we have been able to emulate the other galaxy clusters by calibrating the SAG parameters with a few clusters, demonstrating its robustness and potential for predicting galaxy properties across a wide range of halo masses. This highlights the capability of SAG to provide a computationally efficient alternative to hydrodynamical simulations while maintaining consistency with observationally relevant trends.

        \subsection{Comparison between 7K and 15K resolution simulations}
            \label{subsection: 7K and 15K comparison}

            \begin{figure}
                \includegraphics[scale=0.58]{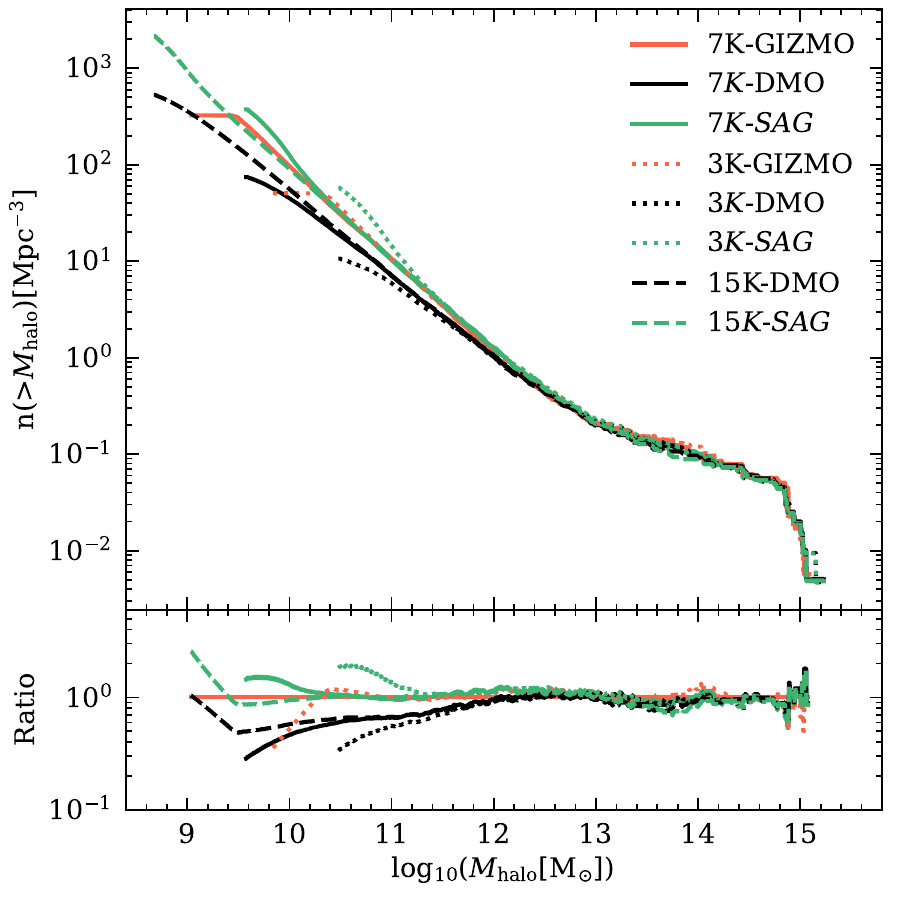}
                \caption{Cumulative halo mass function (CHMF) at $z=0$ is presented for all halo clusters with $M_{\mathrm{halo}} > 10^{14}$ \modot~ in the coincident regions of all simulation types available in this work: 3K-DMO, 7K-DMO, and 15K-DMO (black lines), and 3K-GIZMO and 7K-GIZMO (red lines). Additionally, we present the CHMFs calculated for the SAG model (green lines), which include the no longer identified DM halos hosting orphan galaxies, with their DM mass assigned based on their inferred dynamical properties (as described in Sect. \ref{subsection: sag}).}
                \label{fig: chmf, 7k and 15k, z=0}
            \end{figure}

            Given that the SAM models were calibrated at a specific mass resolution (7K), we study whether this calibration can be extrapolated to mimic the hydrodynamic runs at higher resolutions. We select the coincident DM clusters with $M_{\mathrm{halo}} > 10^{14}, \mathrm{M}_\odot$ at $z=0$, available at all resolutions in GIZMO and SAMs: 3K, 7K, and 15K. The final selection includes approximately 20 DM clusters, a number determined by the availability of 11 regions at 15K-DMO. This selection will also be applied consistently to the subsequent figures presented in this subsection.

            As shown in Fig. \ref{fig: chmf, 7k and 15k, z=0}, the cumulative halo mass function (CHMF) of 7K-GIZMO exhibits a nearly constant power-law slope across most halo mass ranges, reflecting a self-similar distribution of subhalos. A smooth transition in the slope is observed as subhalos of low and intermediate mass ($M_{\mathrm{halo}} < 10^{12}, \mathrm{M}_\odot$) give way to very massive halos, reflecting the hierarchical growth of structure, where smaller halos merge to form larger halos over time. This change in slope is driven by the merging and accretion processes intrinsic to the hierarchical clustering paradigm and is evident in both hydrodynamical and DMO simulations. At higher masses, around $M_{\mathrm{halo}} \sim 10^{15}\, \mathrm{M}_\odot$, an abrupt change in the slope marks the transition from subhalos to halos hosting central galaxies. This abrupt transition is also preserved in the DMO simulations (black lines), consistent with the behaviour observed in GIZMO. In comparison, the CHMFs from DMO simulations (black lines) systematically underestimate the cumulative number of halos relative to GIZMO at all resolutions, as found by \citet{haggar2021}. This behaviour is mirrored by the SAGE catalogues, which are equivalent to the DMO (not shown in Fig. \ref{fig: chmf, 7k and 15k, z=0}. Consequently, SAGE fails to reproduce the GIZMO CHMF, particularly at lower halo masses ($M_{\mathrm{halo}} < 10^{11}\, \mathrm{M}_\odot$). On the other hand, SAG (green lines) shows good agreement with GIZMO at intermediate and high-mass ranges ($M_{\mathrm{halo}} > 10^{11}\, \mathrm{M}_\odot$), but deviates at lower halo masses where SAG significantly overestimates the abundance of halos compared to GIZMO. This overabundance is linked to the treatment of orphan galaxies. SAG includes tidal stripping and dynamical friction, which progressively reduce the mass of orphan galaxies’ halos over time, as described in \citet{cora2018} and \citet{delfino2021}. Thus, the subhalo mass continues to decrease when it disappears from the merger tree due to these effects. In 3K-SAG, the presence of orphans allows the CHMF to follow 3K-GIZMO from high masses down to \(10^{11.3} M_{\odot}\). However, between \(10^{11.3} M_{\odot}\) and \(10^{10.5} M_{\odot}\), the specific physical treatment of orphan galaxies leads to a systematic overabundance of halos compared to 3K-GIZMO. This excess is likely due to an underestimation of the tidal stripping effect on smaller orphan halos, allowing them to survive longer than in hydrodynamical simulations. A notable trend in SAG is that this effect shifts towards lower masses as the simulation resolution increases. While in 3K-SAG the CHMF follows 3K-GIZMO down to \(10^{11.3} M_{\odot}\), in 7K-SAG this agreement extends down to \(10^{10.4} M_{\odot}\) with respect to 7K-GIZMO, covering an additional order of magnitude. Following this trend, an extrapolation to a hypothetical 15K-GIZMO simulation can be made. The separation is expected to follow the resolution scaling factor of 0.903 dex between consecutive resolutions, derived from the number of particles in the simulations. Thus, the separation with 15K-SAG should occur around \(10^{9.5} M_{\odot}\). This suggests that, although the inclusion of orphan galaxies in SAG allows the model to match the hydrodynamical CHMF over a broader mass range, the tidal stripping effects applied to the smallest orphan halos may not be sufficient to remove low-mass halos at the rate predicted by hydrodynamical simulations, leading to their overabundance. Similar effects at low masses have been reported in other studies, such as \citet{haggar2021} and \citet{knebe2017}, highlighting that the inclusion of orphan galaxies can amplify discrepancies between SAMs and hydrodynamical simulations in this regime. Unlike GIZMO, where the CHMF smoothly is extended to lower halo masses at higher resolution, the SAG results show steeper slopes and deviations at lower masses. These results demonstrate the ability of 15K-SAG to extend the CHMF beyond the resolution limits of GIZMO and infer results for not-yet-simulated 15K hydrodynamical resolutions, while emphasizing the need to refine the treatment of orphan galaxies to better align with the smooth behaviour observed in GIZMO simulations.

            \begin{figure}
                \includegraphics[scale=0.58]{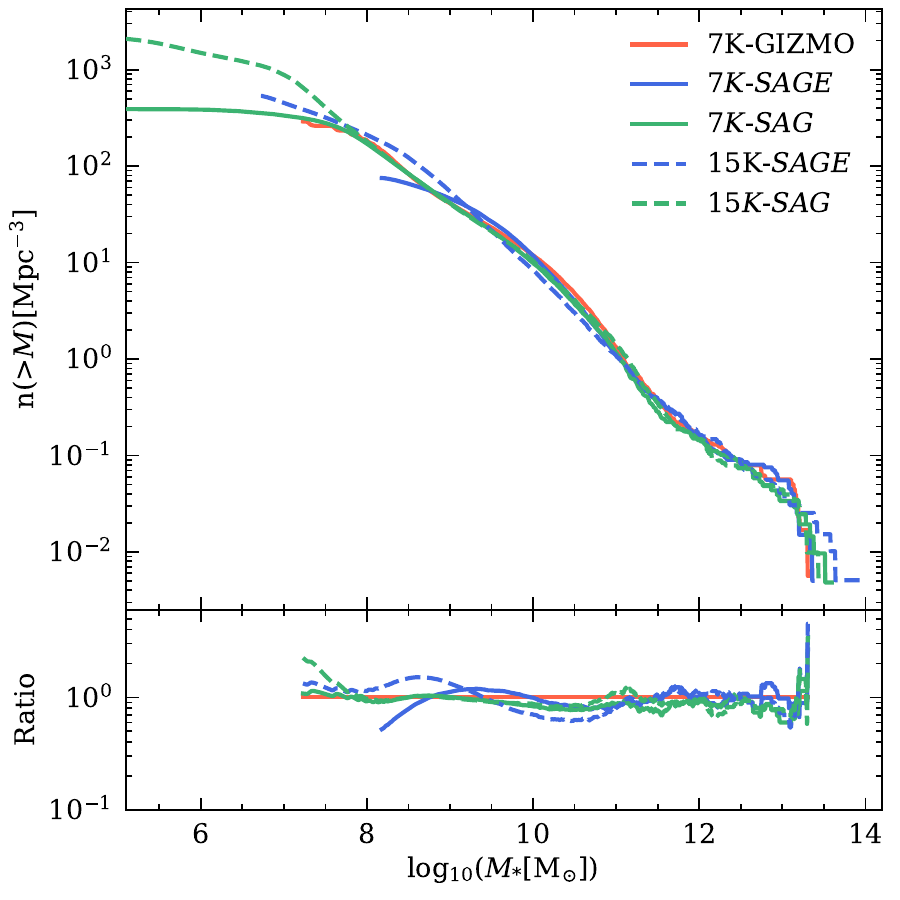}
                \caption{Cumulative stellar mass function (CSMF) at $z=0$ from all halo clusters with $M_{\mathrm{halo}} > 10^{14}$ \modot~ in the coincident regions of all flavours of simulations available in this work. The bottom panel shows the ratio of all the CSMFs compared with that of 7K-GIZMO.}
                \label{fig: csmf, 7k and 15k, z=0}
            \end{figure}
            
            Regarding the galaxy properties, we present the CSMFs of each model for the coincident regions in Fig. \ref{fig: csmf, 7k and 15k, z=0}. The 7K-GIZMO model exhibits a consistent power-law slope in the intermediate stellar mass range ($10^8 \, \mathrm{M}_\odot < M_* < 10^{12} \, \mathrm{M}_\odot$), with transitions at low stellar masses ($M_* \lesssim 10^8 \, \mathrm{M}_\odot$) due to resolution effects and at high stellar masses ($M_* \gtrsim 10^{13} \, \mathrm{M}_\odot$) dominated by central galaxies. The 15K-SAGE model (dashed blue line) fails to reproduce the behaviour of GIZMO. It exhibits an inflection point at $M_* = 10^{9.5} \, \mathrm{M}_\odot$, where it overpredicts the number of galaxies below this mass and underpredicts them above it. This behaviour reflects a sensitivity to resolution changes, as SAGE does not handle these scales consistently when calibrated at 7K resolution. Consequently, the results from 15K-SAGE highlight the inability to extrapolate its calibration to higher resolutions without significant recalibration. In contrast, the 15K-SAG model (dashed green line) successfully extends the results of 7K-SAG to higher resolutions. It maintains a good agreement with 7K-GIZMO across the entire stellar mass range and even extends the CSMF down to $\mathrm{M}_* = 10^7 \, \mathrm{M}_\odot$, an order of magnitude lower than 7K-GIZMO, without flattening. This demonstrates the robustness of the SAG calibration and its ability to consistently capture galaxy properties at higher resolutions. Furthermore, the consistent power-law slope at intermediate masses and the preserved transitions at low and high masses underscore the capability of 15K-SAG to emulate hydrodynamical results effectively while extending the predictive power of DMO simulations combined with SAG.

            \begin{figure}
                \includegraphics[scale=0.59]{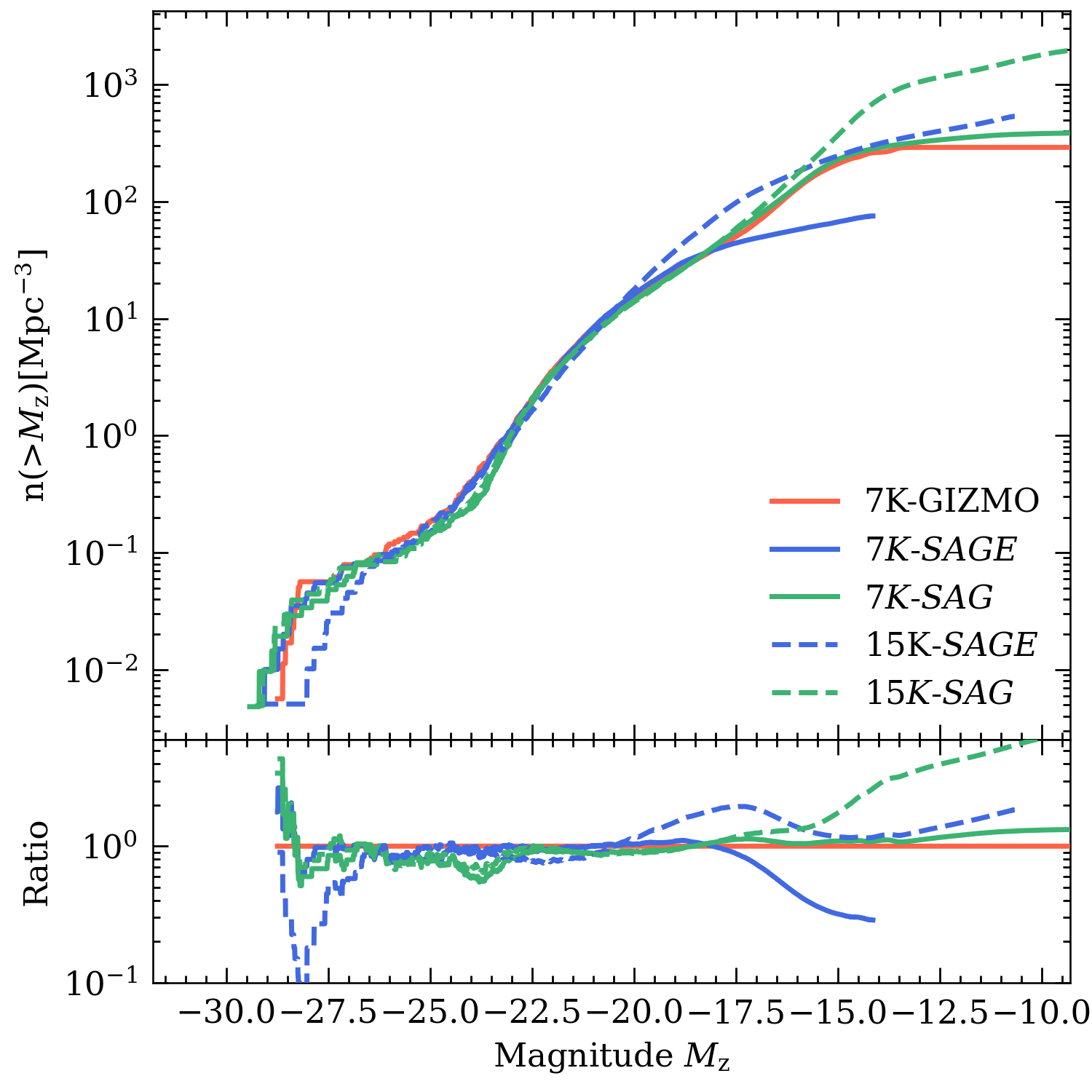}
                \caption{Cumulative luminosity function (CLF) for the absolute magnitude in the $z$-band at $z=0$, derived from all halo clusters with $M_{\mathrm{halo}} > 10^{14} M_\odot$ in the coincident regions of all simulation types used in this work, where the $z$-band was one of the bands employed in the calibration of the 7K-SAM. The bottom panel shows the ratio of all the CLFs compared with that of 7K-GIZMO.}
                \label{fig: clf, 7k and 15k, z=0}
            \end{figure}
            
            In Fig. \ref{fig: clf, 7k and 15k, z=0}, we present the CLF for the $z$-band for all models, derived from the coincident regions. As expected, The 7K-GIZMO model shows a smooth power-law slope in the intermediate luminosity range ($-20 \lesssim M_{\mathrm{z}} \lesssim -27$), with transitions at low luminosities ($M_{\mathrm{z}} \gtrsim -20$) due to resolution effects and at high luminosities ($M_{\mathrm{z}} \lesssim -27$) dominated by central galaxies. The 15K-SAGE model (dashed blue line) deviates significantly, exhibiting an inflection at $M_{\mathrm{z}} \sim -20$ where it overpredicts galaxies at brighter magnitudes and underpredicts them at fainter magnitudes, highlighting its sensitivity to resolution changes and the need for recalibration. Conversely, the 15K-SAG model (dashed green line) closely matches 7K-GIZMO across the luminosity range, extending the CLF smoothly down to $M_{\mathrm{z}} \sim -14$, two magnitudes lower than 7K-GIZMO.

            \begin{figure}
                \includegraphics[scale=0.59]{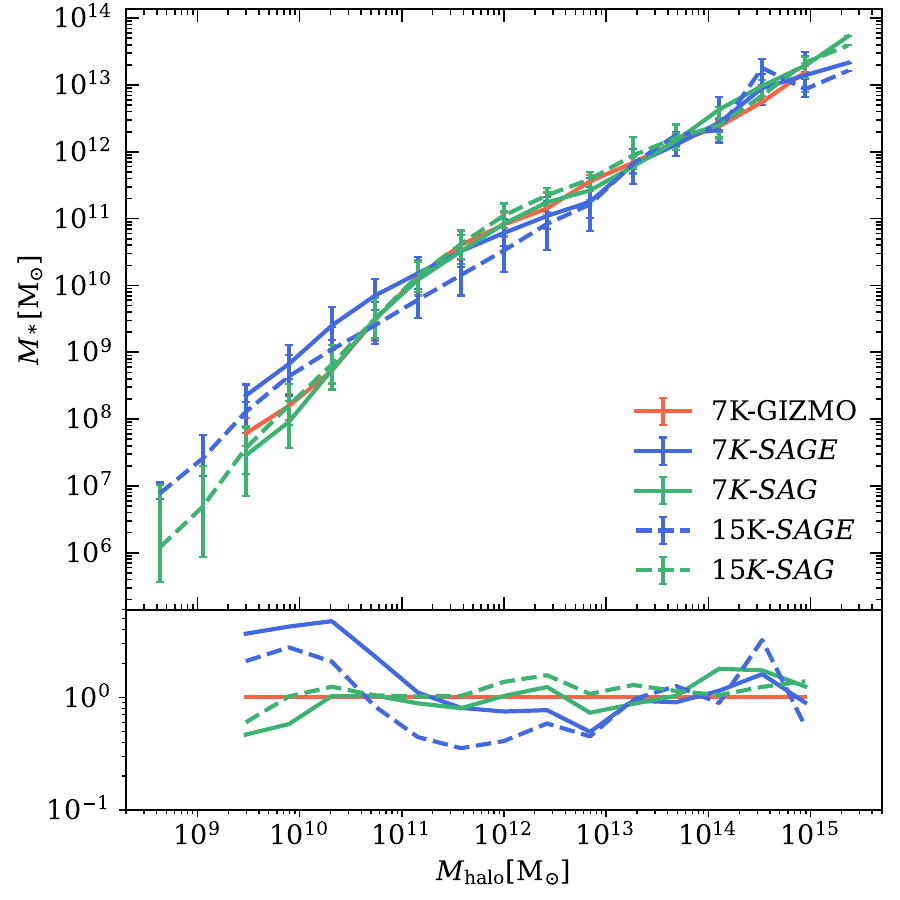}
                \caption{Stellar mass - Halo mass relationship at $z=0$ from all halo clusters with $M_{\mathrm{halo}} > 10^{14}$ \modot~ in the coincident regions of all flavours of simulations available in this work. The lines represent the median and the error bars represent the 25th and 75th percentiles.}
                \label{fig: smass vs hmass, 7k and 15k, z=0}
            \end{figure}

            In Fig. \ref{fig: smass vs hmass, 7k and 15k, z=0}, we show the stellar mass–halo mass ($M_{\mathrm{halo}} - M_{\mathrm{*}}$) relation at $z=0$ for the coincident regions across all simulation types. The 7K-GIZMO model presents a smooth increase in stellar mass as a function of halo mass, with no significant deviations or inflection points, serving as the reference. 7K-SAGE deviates significantly from 7K-GIZMO. At $M_{\mathrm{halo}} < 10^{11} \, \mathrm{M}_\odot$, 7K-SAGE overestimates the stellar mass, while at $M_{\mathrm{halo}} > 10^{12} \, \mathrm{M}_\odot$, it underestimates the stellar mass, an issue that persists and is exacerbated in 15K-SAGE. The latter shows additional inconsistencies, such as deviations across the entire halo mass range, demonstrating its sensitivity to resolution changes and limiting its extrapolation capabilities. In contrast, 7K-SAG follows 7K-GIZMO closely across nearly all mass ranges, and its extrapolation to 15K-SAG maintains this agreement. Notably, 15K-SAG extends the relation smoothly to lower stellar masses while preserving the trend observed in 7K-GIZMO, reflecting the robustness of SAG's calibration. As expected, these results emphasize the capability of SAG to predict galaxy properties consistently across multiple resolutions, aligning with observational trends.

            The 15K resolution results highlight the robustness of SAG compared to SAGE when extrapolating beyond the 7K calibration. While SAGE struggles to maintain consistency with hydrodynamical simulations, SAG not only preserves the trends of 7K-GIZMO but also extends predictions to lower stellar masses and fainter magnitudes than hydrodynamical simulations. This highlights the importance of orphan galaxy modelling and demonstrates SAG's capability as a computationally efficient tool to bridge the gap between DMO and hydrodynamical simulations, enabling reliable predictions for galaxy properties across a wide range of halo masses and resolutions. Furthermore, its robustness at higher resolutions makes SAG well-suited for interpreting data from future deep galaxy cluster surveys.

        \subsection{Prediction of galaxy properties}
            \label{subsection: prediction of galaxy properties}

            The main objective of this work is to establish a method capable of predicting the behaviour of a large sample of hydrodynamical simulations using calibrated SAMs. Specifically, we aim to predict galaxy properties at any redshift and extend these predictions to different resolutions by leveraging DMO simulations. In this Section, we analyse galaxy properties that were not included as constraints during the calibration of the models.

            \begin{figure}
                \includegraphics[scale=0.59]{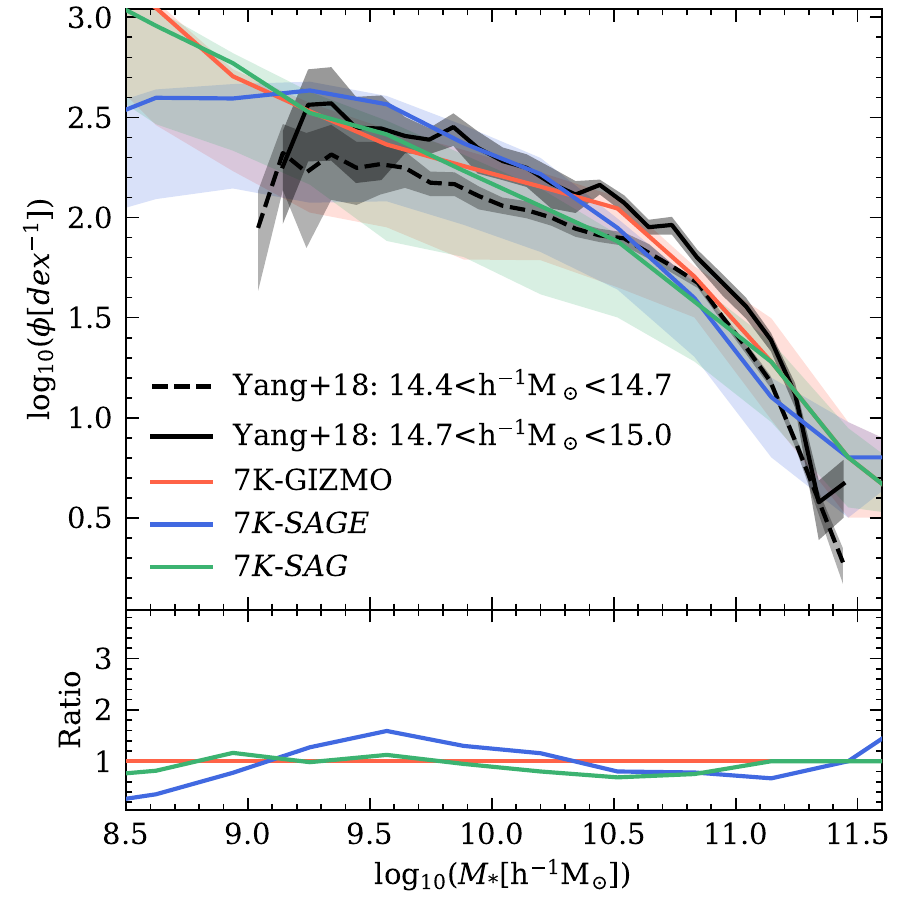}
                \caption{Median satellite stellar mass function ($\mathrm{<SSMF>}$) in different models and observations. We show the median values (solid lines) with 1$\sigma$ error bars (shaded regions) in each stellar mass bin, obtained from the SSMFs of the clusters. The observational results within two particular halo mass ranges are shown in black lines from \citet{yang2018}, while the $\mathrm{<SSMF>}$ from the models 7K-GIZMO, 7K-SAGE, and 7K-SAG are displayed in red, blue, and green, respectively. This plot corresponds to Fig. 8 in \citet{cui2022:gizmo} but with the results from the GADGET-X and GIZMO models at 3K resolution removed.}
                \label{fig: <ssmf>, 7k and observations, z=0}
            \end{figure}

            Fig. \ref{fig: <ssmf>, 7k and observations, z=0} presents the median satellite stellar mass function ($\mathrm{<SSMF>}$) for our simulations and observational data. The observational $\mathrm{<SSMF>}$ of galaxy clusters from \citet{yang2018}, originally derived from \citet{yang2012} and based on the Sloan Digital Sky Survey \citep[SDSS;][]{york2000} at $0.01 < z < 0.12$, is compared with the $\mathrm{<SSMF>}$ from our models—7K-GIZMO, 7K-SAGE, and 7K-SAG. The 3K-GIZMO simulations analysed in this work were calibrated to match the observed \(\mathrm{<SSMF>}\) in \citet{cui2022:gizmo}, ensuring that their galaxy populations closely followed observational constraints. The same calibration procedure was applied to 7K-GIZMO, maintaining consistency across different resolutions. A key observation from this figure is that 7K-GIZMO exhibits the same level of agreement with the observed \(\mathrm{<SSMF>}\) as 3K-GIZMO in \citet{cui2022:gizmo}, further validating the stability of the calibration and reinforcing the reliability of hydrodynamical simulations in capturing the properties of satellite galaxies in clusters. The 7K-SAMs also reproduce the general trend, exhibiting the same characteristic steepness as the observations, and systematically overpredict the number of massive satellite galaxies at high stellar masses (\( M_{\mathrm{*}} > 10^{11} M_{\odot} \)), which is identical with what is observed in 7K-GIZMO and also seen in 3K-GIZMO from \citet{cui2022:gizmo}. This deviation may be partially attributed to an incompleteness in the cluster sample from \citet{yang2018}, which could lead to an underestimation of the most massive satellite galaxies in the observational dataset. Moreover, 7K-SAGE tends to overpredict the abundance of satellites in the stellar mass range \( 10^{9} M_{\odot} < M_{\mathrm{*}} < 10^{10.5} M_{\odot} \). This overestimation is likely a consequence of our calibration process, which adjusts 7K-SAGE to match 7K-GIZMO. The limited resolution of the simulation constrains the number of galaxies with \( M_{\mathrm{*}} < 10^{9} M_{\odot} \) that can be resolved in 7K-SAGE, which may introduce biases in the calibration at the low-mass end. Despite these small deviations, the overall agreement between 7K-SAMs, 7K-GIZMO, and the observational $\mathrm{<SSMF>}$ confirms that the semi-analytic models accurately reproduce the satellite galaxy populations and their connection to the underlying dark matter distribution.

            \begin{figure*}
                \includegraphics[scale=0.59]{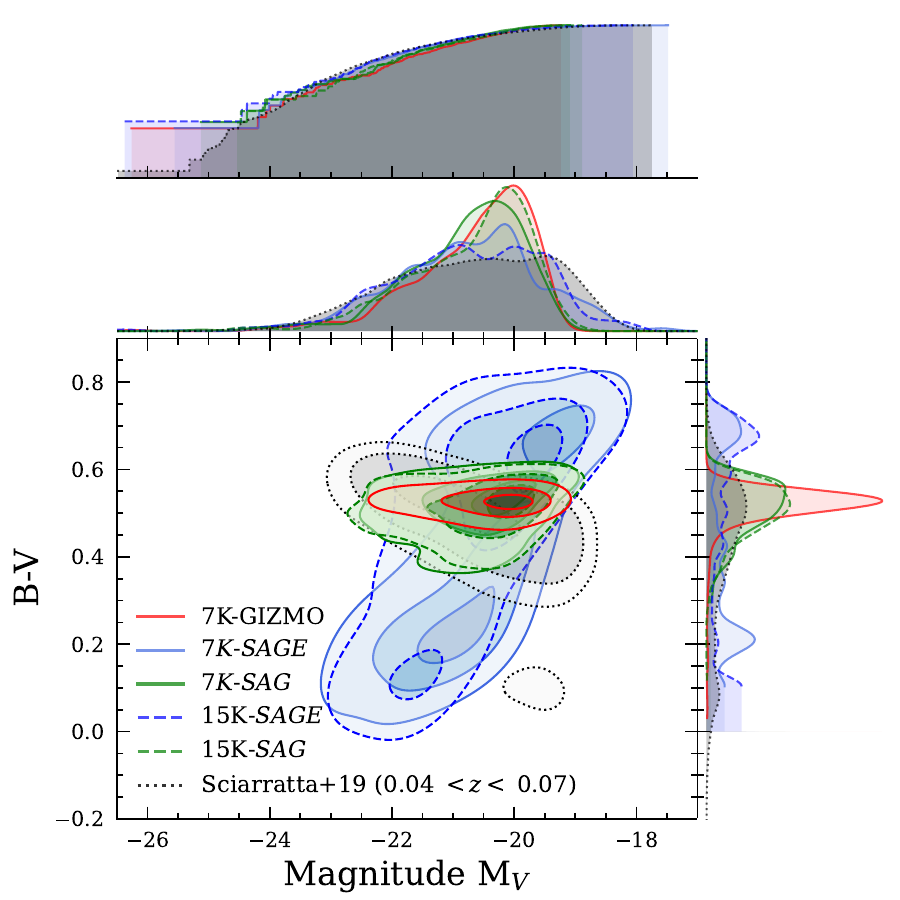}
                \includegraphics[scale=0.59]{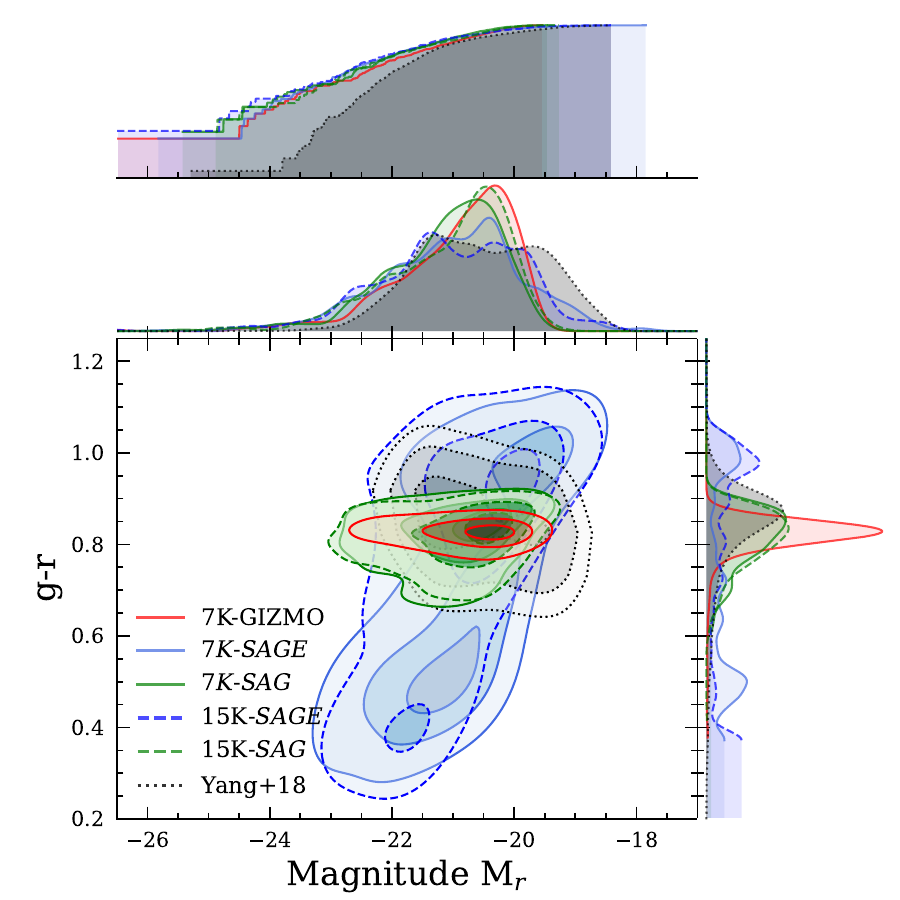}
                \caption{Left panels: the Colour-Magnitude Diagram for B-V bands (biggest panel), normalised cumulative magnitude function (upper panel), density histogram for x-axis and y-axis (second-top and left panel, respectively) for 7K-GIZMO, 7K-SAGE, 7K-SAG, 15K-SAGE and 15K-SAG (see labels for colour and styles). The black dotted contours correspond to observations of nearby galaxy clusters obtained in \citet{sciarratta2019}. Right panels: as left panels but with the bands g-r. In addition, the observations of nearby galaxy clusters obtained in \citet{yang2018} are shown. The contour levels of each biggest panels correspond to 0.5, 1 and 2 sigma of the data.}
                \label{fig: CMDs [B-V, g-r], 7k and 15k, z=0}
            \end{figure*}

            In Fig. \ref{fig: CMDs [B-V, g-r], 7k and 15k, z=0}, we show the B-V (left panel) and g-r (right panel) colour-magnitude diagrams (CMDs) of the 7K-GIZMO and both SAMs at 7K and 15K resolutions. As we mentioned, all the absolute magnitudes were calculated using STARDUST stellar population synthesis algorithm. Additionally, in the B-V vs. $M_v$ CMD we include observational data of nearby galaxy clusters at $0.04 < z < 0.07$ extracted from \citet{sciarratta2019}. These clusters were originally studied in \citet{cariddi2018} from the Omega-WINGS survey and include galaxies with $M_{\mathrm{*}} > 10^{10} \, \mathrm{M}_\odot$. Similarly, in the g-r vs. $M_r$ CMD, we include observations of galaxy clusters from SDSS at $0.01 < z < 0.12$, which also include galaxies with $M_{\mathrm{*}} > 10^{10} \, \mathrm{M}_\odot$ \citep{yang2018}. For consistency, the same mass cut of $M_{\mathrm{*}} > 10^{10} \, \mathrm{M}_\odot$ was applied to our galaxy sample in this analysis. The upper middle panel of each CMD show the normalised density histograms of each of their bands on those axes. Finally, in the uppermost panel we show the normalised cumulative function (NCF) of the absolute magnitudes in the \rm{V} (left) and \rm{r} (right) bands for each sample. The 7K-GIZMO model (red solid line) shows smooth CMD contours that align well with the observational data. The colour-magnitude distribution exhibits a smooth and continuous gradient, with no evident bimodality, as it is dominated by the red sequence—consistent with observations of galaxy clusters from \citet{baldry2006}—as expected in high-density environments. This reflects the ability of hydrodynamical processes to regulate star formation and feedback, leading to a well-defined population of quenched galaxies. The NCFs of 7K-GIZMO closely match the observations, particularly in the V and r bands, confirming the reliability of the hydrodynamical model at capturing galaxy magnitudes. 
            
            For the SAGE model at 7K and 15K, we can see in both CMDs that the contours are slightly displaced from each other. As expected, This displacement highlights the sensitivity of this model to resolution changes. Furthermore, SAGE exhibits a bimodal colour distribution even though these have very similar NCFs in absolute magnitude. This similarity between the NCF of 7K-GIZMO and SAGE indicates that the SAGE calibrations in the z and U bands are performed correctly and can be extrapolated to other bands. However, the situation seems to be that, while the NCFs work fine statistically, the combination of these properties, such as colours for individual galaxies, fails to capture the expected behaviour. Independent of this similarity, we see that the galaxies of 7K-SAGE/15K-SAGE are separated into two populations, some galaxies being redder and others bluer than those of 7K-GIZMO. This bimodality suggests that the SAGE galaxies had a different star formation history than those of 7K-GIZMO despite forcing a calibration of galaxy properties at different redshifts. This discrepancy arises from how SAGE treats subhalos that are no longer resolved in the merger trees. When this occurs, the associated galaxy is no longer tracked, and its properties remain frozen until it merges with a central galaxy. As a result, galaxies that should experience quenching due to environmental effects, such as gas stripping and AGN feedback, retain their pre-disruption properties, leading to a population that remains bluer. Furthermore, since these unresolved subhalos host blue galaxies that do not undergo quenching, their mergers with more massive galaxies contribute to making the brightest galaxies in SAGE bluer than expected. This continuous accretion of star-forming satellites, which remain blue due to the lack of environmental quenching, enhances the blue colours of the most luminous galaxies. In contrast, galaxies in well-tracked subhalos continue evolving, undergoing quenching and progressively becoming redder. A similar effect was observed in \citet{knebe2017}, where SAGE was applied to the DMO MDPL2 simulation \citep{klypin2016}. Given this, the SAGE model deviates strongly from the CMDs of GIZMO and the observations by \citet{sciarratta2019} and \citet{yang2018}. In contrast, the SAG-7K and SAG-15K are almost coincident with each other, and follow closely the GIZMO CMD contours. Moreover, the distribution centres in both B-V and g-r CMDs of SAG, GIZMO and the observations are coincident. Although these distribution centres coincide, SAG exhibits a broader spread compared to 7K-GIZMO, which may reflect statistical variations in the colour distribution of galaxies. This broader spread, however, results in a better representation of the observations than GIZMO. Finally, as in SAGE, SAG and 7K-GIZMO have very similar NCFs in absolute magnitude V and r, so the 7K-SAG calibration can be extrapolated to other magnitudes. Moreover, SAG gives us a hint that we can extrapolate a calibration of galaxy property functions to other functions that are not considered in the calibration.

            \begin{figure}
                \includegraphics[scale=0.59]{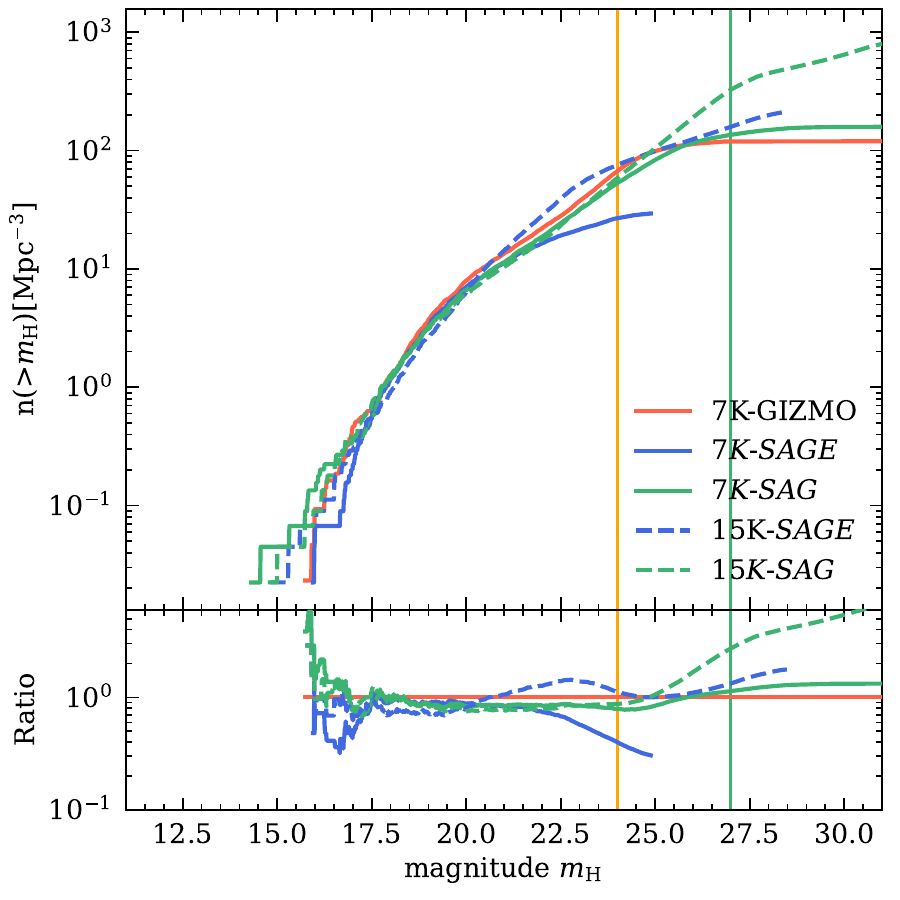}
                \caption{Cumulative luminosity function for apparent magnitude in $H$-band at $z=1$ of the satellite galaxies from all halo clusters with $M_{\mathrm{halo}} > 10^{14}$ \modot~ in the coincident regions of all flavours of simulations available in this work. The bottom panel shows the ratio compared to the 7K-GIZMO simulations. The vertical orange line represents the observational limit for {\sc Euclid}. The vertical green line represents the beginning of the loss of the power-law slope for 15K-SAG.}
                \label{fig: clf-apparent, H band, 7k and 15k, z=1}
            \end{figure}

            On the other hand, future deep surveys of galaxy clusters will be able to reach very low apparent magnitudes for faint galaxies at high redshifts. In the case of {\sc Euclid} \citep{laureijs2011:euclid}, for example, it will be able to reach magnitude 24 in the $H$-band ($m_H=24$) at $z=1$ \citep{jimenez2024}. To ease the comparison to the results on the Mock {\sc Euclid} catalogue presented in \citet{adam2019}, \citet{jimenez2024} computed the luminosity functions for apparent magnitudes in the $H$-band for the galaxy cluster catalogues of {\sc The Three Hundred} project. In their work, the same galaxy clusters from the five regions used in our calibration for both the GIZMO hydrodynamics simulations at 3K and 7K resolutions were included. In Fig. \ref{fig: clf-apparent, H band, 7k and 15k, z=1} we present the cumulative luminosity function in the $H$-band for apparent magnitudes at $z=1$ of the satellite galaxies from the coincident galaxy clusters between 7K-GIZMO and SAMs (SAGE and SAG) at the 7K and 15K resolution of our work which has a content equivalent to those found in Fig. 1 of \citet{jimenez2024}. We observe that the luminosity function of 7K-GIZMO reaches a maximum magnitude of 24, the minimum limit on {\sc Euclid} magnitude (vertical orange line). For SAGE, we can reach the Euclid limit with only the 15K-DMO simulations, which is not congruent with the 7K-GIZMO faint galaxies. On the other hand, for SAG, we can reach the same Euclid magnitude limit (and therefore 7K-GIZMO as well) with the 7K-SAG simulations and continue extrapolating the results to even fainter galaxies with 15K-SAG. With 15K-SAG we can obtain galaxies with faint apparent magnitudes of up to $m_{\rm{H}} = 27$ (vertical green line) without the accumulated function being substantially flattened. This limit of $m_{\rm{H}} = 27$ reached will open the possibility of exploring dwarf galaxies at high redshift and studying their evolution in the densest existing environment. However, 15K-SAG allows us to probe even fainter apparent magnitudes, m$_H \sim 31$, without the cumulative function flattening completely, as occurs with 7K-GIZMO at m$_H = 24$. In the apparent magnitude range of 27 to 31, the slope of the power-law function for 15K-SAG decreases, which we also observed in the CSMF of Fig. \ref{fig: csmf, 7k and 15k, z=0} and the CLF at z-band of Fig. \ref{fig: clf, 7k and 15k, z=0}. However, the nature and properties of these faint objects remain uncertain, as the SAG model has not been extensively explored in this resolution regime. This decrease in the slope for small galaxies could be due to incompleteness provided by the modelling process of the galaxies (mainly dominated by orphan galaxies) of SAG, or because even though these galaxies are well modelled we find physical processes dependent on the stellar mass content of the objects. The latter suggests that the decrease in the slope could be linked to a transition in the properties or evolutionary processes of low-mass objects, such as dwarf galaxies or other faint systems. This transition remains uncertain and highlights the need for further exploration in future studies. In any case, with the 15K-SAG emulator we will be able to contrast upcoming Euclid catalogues with valid synthetic data up to $m_{\rm{H}} \sim 24$, with the aim of study and predict the properties of dwarf galaxies in high density environments.

    \section{Summary and conclusions}
        \label{section: conclusions}

        In this paper, we present a novel method to emulate galaxy properties from hydrodynamical simulations of galaxy clusters using semi-analytical models (SAMs). The emulators were developed and applied to cosmological DMO simulations from {\sc The Three Hundred} project. The SAM parameters were calibrated using Particle Swarm Optimisation, incorporating constraints on stellar mass and luminosities across multiple bands and redshifts ($0 \leq z \leq 1$). This calibration was performed using a few regions, comparing DMO clusters with their hydrodynamical counterparts simulated with GIZMO at 7K resolution.

        Our results confirm that these calibrated SAMs accurately replicate the galaxy properties observed in hydrodynamical simulations of galaxy clusters. Applying these emulators to higher-resolution DMO simulations allows us to not only match the accuracy of hydrodynamical simulations but also surpass their resolution limits, all while significantly reducing computational costs. The key results from these calibrated SAMs are as follows:

        \begin{itemize}
            \item The 7K-SAGE emulator, while achieving a reasonable fit to 7K-GIZMO for the constraints used, fails to reach the same lower limits in stellar masses and luminosities. Furthermore, SAGE is highly sensitive to resolution changes and cannot extrapolate its calibration from 7K (used for calibration) to 15K resolutions. Although 7K-SAGE correctly reproduces stellar masses and luminosities in the $z$- and $U$-bands for 145 additional regions beyond those used in calibration, it struggles to predict other galaxy properties, such as the $M_{\mathrm{halo}} - M_{\mathrm{*}}$ relation and galaxy colours, both at 7K and 15K resolutions.
        
            \item The 7K-SAG emulator closely reproduces 7K-GIZMO results with high precision, reaching lower stellar masses and luminosities than 7K-GIZMO. Importantly, SAG is robust to resolution changes: it extrapolates galaxy properties not included as constraints during calibration and achieves consistent results for the 145 regions not used in calibration. Furthermore, the 7K-SAG calibration extends effectively to the 15K resolution, preserving the accuracy and predictive capability observed at 7K. This success can be attributed to SAG's treatment of orphan galaxies, which ensures a more accurate representation of low-mass galaxies and their evolution in dense environments.
        \end{itemize}
        
        Our results demonstrate that SAG outperforms SAGE in this regard, offering a robust and scalable solution for emulating galaxy properties across a wide range of resolutions. With 15K-SAG, we achieve faint galaxy apparent magnitudes up to $m_{\mathrm{H}} = 27$, three magnitudes fainter than 7K-GIZMO, and stellar masses as low as $10^7 \, \mathrm{M}_\odot$. In the range of $m_{\mathrm{H}} = 27$ to 31, SAG begins to exhibit a decrease in the power-law slope in cumulative functions, possibly due to the modelling of orphan galaxies or physical transitions from dwarf galaxies to globular clusters. While these effects require further study, SAG enables detailed exploration of dwarf galaxies in high-density environments at unprecedented depths.
        
        This methodology bridges the gap between DMO simulations and hydrodynamical models, leveraging the computational efficiency of SAMs while achieving comparable results. The CMD predictions generated by SAG, consistent with GIZMO at 7K and extendable to higher resolutions, provide a robust framework for studying galaxy evolution. By producing catalogs compatible with hydrodynamical simulations and enabling predictions at unprecedented resolutions, SAG offers a powerful tool for exploring dwarf galaxies and their properties in dense cluster environments. These predictions are particularly valuable for interpreting data from upcoming deep surveys, such as {\sc Euclid}, LSST, and {\sc 4MOST}, shedding light on the formation and evolution of galaxies in the densest regions of the universe.

    \begin{acknowledgements}
    This work has been made possible by {\sc The Three Hundred} (https://the300-project.org) collaboration.  JSG acknowledges funding from Predoctoral contract ‘Formaci\'on de Personal Investigador’ from the Universidad Aut\'onoma de Madrid (FPI-UAM, 2021). G.Y., JSG, AJM and  W.C. would like to thank Ministerio de Ciencia e Innovación (MICINN) for financial support under the project grant PID2021-122603NB-C21. WC is further supported by Atracci\'{o}n de Talento fellowship no. 2020-T1/TIC19882 granted by the Comunidad de Madrid and by the Consolidación Investigadora no. CNS2024-154838 granted by the Agencia Estatal de Investigación (AEI) in Spain. He also thanks the science research grants from the China Manned Space Project and the ERC: HORIZON-TMA-MSCA-SE for supporting the LACEGAL-III project with grant number 101086388. TH acknowledge Consejo Nacional de Investigaciones Cient\'ificas y Técnicas (CONICET), Argentina, for their supporting fellowships. SAC acknowledge funding from CONICET (PIP KE4-11220200102876CO), and support from the Universidad Nacional de La Plata (11/G183), Argentina. The authors acknowledge The Red Española de Supercomputaci\'on for granting computing time for running the hydrodynamical and DMO simulations of {\sc The Three Hundred} galaxy cluster project in the Marenostrum supercomputer at the Barcelona Supercomputing Center and Cibeles Supercomputers through various of RES grants. The HD hydrodynamic simulations (7K and 15K runs) were performed also on the DIaL3 -- DiRAC Data Intensive service at the University of Leicester through the RAC15 grant: Seedcorn/ACTP317, and on the Niagara supercomputer at the SciNet HPC Consortium. DIaL3 is managed by the University of Leicester Research Computing Service on behalf of the STFC DiRAC HPC Facility (www.dirac.ac.uk). The DiRAC service at Leicester was funded by BEIS (ST/K000373/1), UKRI, STFC capital funding, and STFC operations grants (ST/K0003259/1). DiRAC is part of the UKRI Digital Research Infrastructure. SciNet \citep{Loken_2010} is funded by Innovation, Science and Economic Development Canada; the Digital Research Alliance of Canada; the Ontario Research Fund: Research Excellence; and the University of Toronto. This work made also use of the local computer facilities at the Universidad Aut\'onoma de Madrid. We would also like to thank the anonymous referee for his/her  careful reading of the manuscript and for his/her comments that have contributed to  greatly improved the quality of the paper.
    \end{acknowledgements}

    \section*{Data Availability}
    The results shown in this work use data from \tth~ galaxy clusters sample. These data are available upon request following the guidelines of \tth~ collaboration, at https://the300.ft.uam.es. The data specifically shown in this paper will be shared upon request with the authors.

    \bibliography{references}

    \begin{appendix}
        \section{SAM parameters}
        \label{appendix - seccion: sam parameters}

            In Section \ref{section: calibration}, we described the calibration process of the SAGE and SAG semi-analytical models at 7K resolution. Below, we provide a brief description of the key parameters used in the calibration of these SAMs.

            \subsection{Free parameters of the SAGE model}
                \label{appendix - subsection: sage parameters}

                The optimized parameters in the SAGE calibration are strongly related to the constraints used, and here we describe the nine most important of these:
                
                \begin{itemize}
                    \item \SfrEfficiency:
                        SAGE assume that only cold disc gas can form stars, either quiescently or in a burst. Above a critical surface density threshold, the cold gas collapse and form stars \citep{kennicutt1998}. SAGE convert this critical surface density into a critical mass by assuming the cold gas mass to be evenly distributed over the disc following the work of \cite{kauffmann1996} and complemented with \cite{mo1998} and \citep{bullock2001}. When the mass of cold gas in a galaxy is greater than this critical value SAGE calculate the star formation rate from a Kennicutt–Schmidt-type relation \citep{kennicutt1998}, Therefore, a fraction $\alpha_{\mathrm{SF}}$ of gas above the threshold at that relation is converted into stars in a disc dynamical time $t_{\mathrm{dym},\, \mathrm{disc}} = r_{\mathrm{disc}} / V_{\mathrm{vir}}$.
                        
                    \item \FeedbackReheatingEpsilon:
                        As star formation proceeds, newly formed very massive stars rapidly complete their evolution and end their life as supernovae. Supernovae play an important role in the life-cycle of a galaxy, injecting metals, gas and energy into the surrounding interstellar medium, reheating cold disc gas and possibly ejecting gas even from the surrounding halo. SAGE assume that supernova winds remove cold gas from the disc, which in turn acts to suppresses star formation. The rate at which disc cold gas is reheated by supernovae from the galaxy is suggested by observations of \citet{martin1999}. The proportionality constant, $\epsilon_{\mathrm{disc}}$, is typically referred to as a mass-loading factor.
                        
                    \item \FeedbackEjectionEfficiency:
                        The energy released by supernovae during the star formation episode can be approximated as $\dot{E}_{\mathrm{SN}} = 0.5 V^2_{\mathrm{SN}} \epsilon_{\mathrm{halo}} \dot{m}_*$, 
                        where $0.5 V^2_{\mathrm{SN}}$ is the mean energy in supernova ejecta per unit mass of stars formed, and $\epsilon_{\mathrm{halo}}$ parametrizes the efficiency with which this energy is able to reheat disc gas.
                        
                    \item \ReIncorporationFactor:
                        In a dynamically evolving universe, gas that is ejected may not stay ejected forever. A better match with the data can be obtained when SAGE allows the reincorporation rate (fraction of the ejected material returned to the hot halo) to increase for the more massive halos, and limit it to zero for the very lowest mass halos \citep[suggested by][]{mutch2013}. SAGE assumes that, at each time-step, a fraction of the ejected gas is reincorporated into the hot halo, and $\kappa_\mathrm{reinc}$ parameterizes how efficient this idealized process actually is.
                    
                    \item \RadioModeEfficiency:
                        The hot gas accretes onto the central black hole at a rate approximated using the Bondi–Hoyle formula \citep{bondi1952} (and updated in \citet{croton2006} using the so-called “maximal cooling flow” model of \cite{nulsen&fabian2000}). The “radio mode efficiency” parameter, $\kappa_\mathrm{R}$, was introduced to correct for approximations in the black hole accretion model and to modulate the strength of radio mode feedback.
                        
                    \item \QuasarModeEfficiency:
                        SAGE adopt a simple phenomenological model that is consistent with the quasar mode feedback narrative \citep[][and references therein]{lynden-bell1969, novikov&thorne1973, costa2014, stevens2015}. When a merger or disc instability occurs and the black hole has undergone some form of rapid accretion, we assume a quasar wind follows with luminosity $L_{\mathrm{BH, Q}} = \eta\, \dot{m}_{\mathrm{BH, Q}}\, c^2$. This is used to calculate the total energy contained in the quasar wind. $\kappa_{\mathrm{Q}}$ parametrizes the efficiency with which the wind influences the surrounding gas as it escapes the galaxy and halo.
                        
                    \item \BlackHoleGrowthRate:
                        In most simulations of galaxy formation, quasars are triggered by mergers or from some form of instability in the disc. To model the effect of mergers on black hole growth SAGE follows the work of \cite{Kauffmann&Haehnelt2000}. The constant $f_{\mathrm{BH}}$ controls the fraction of cold gas accreted by a black hole and is modulated by the satellite-to-central galaxy merger mass ratio.
                        
                    \item \ThreshMajorMerger:
                        Once the occurrence of a galaxy–galaxy merger has been identified, SAGE checks the satellite-to-central baryonic mass ratio. If the ratio is above a threshold $f_{\mathrm{major}}$, SAGE says the merger is major. In a major merger the discs of both galaxies are destroyed and all stars are combined to form a spheroid. Otherwise the merger is minor, and only the satellite stars are added to the central galaxy bulge. Furthermore, any cold gas present in either system can lead to a starburst.
                    
                    \item \ThresholdSatDisruption:
                        SAGE models the evolution of satellite galaxies without explicitly tracking their orbits after subhalo disruption. Instead, it determines their fate based on estimated merger times, which simplifies their treatment compared to other models. Hot-halo stripping happens in proportion to the DM subhalo stripping. Any hot gas present in the subhalo is allowed to cool onto the satellite. Upon infall, a merger time is calculated for the satellite. SAGE takes this as the average merger time expected for systems of similar properties. Then, SAGE follow the satellite with time and measure the ratio of subhalo-to-baryonic mass. When this ratio falls below a critical threshold, \ThresholdSatDisruption, SAGE compares its current survival time with the average time determined at infall. If the subhalo has survived longer than average, then SAGE says that the subhalo/satellite system was more bound than average and merge it with the central. On the other hand, if the subhalo/satellite mass ratio has fallen below the threshold sooner than average, then SAGE argues that the system was instead loosely bound and more susceptible to disruption. In this case, SAGE adds the satellite stars to a new intra-cluster stars component, and any remaining gas goes to the parent hot halo.
                        
                \end{itemize}

            \subsection{Free parameters of the SAG model}
                \label{appendix - subsection: sag parameters}

                The optimized parameters in the SAG calibration are similar in essence to used SAGE free parameters, and a description of nine most important of these is:
                
                \begin{itemize}
                    \item \Alpha:
                    This parameter regulates the star formation efficiency, considering the cold gas available in the galactic disc. The conversion of cold gas into stars depends on both the cold gas density and the dynamical time of the disc. This implementation is inspired by the formulation of \citet{springel2001} and \citet{delucia2004}, but adapted to incorporate the effects of the environment and halo properties, as described in \citet{ruiz2015:1°pso+sam} and \citet{cora2018}. Its calibration is essential to reproduce key observables such as the luminosity function and the stellar mass–halo mass relation at different redshifts, ensuring that the model remains robust across diverse galactic environments.
                    \item \Epsilon:
                    This free parameter is associated to disc and bulge. \Epsilon$_\mathrm{disc}$: This controls the amount of cold gas reheated by the energy released by SNe generated from quiescent SF that occurred in the disc. \Epsilon$_\mathrm{bulge}$: This controls the amount of bulge cold gas reheated by SNe formed in the bulge when a starburst is triggered. Since the cold gas transferred from the disc to the bulge is gradually consumed, it can also be affected by SNe feedback.
                    \item \EpsilonEjec:
                    The reheated gas is transferred from the cold to the hot phase, subsequently returning to the cold phase through gas cooling taking place in both central and satellite galaxies. However, to avoid an excess of stellar mass in low-mass galaxies at high redshifts, some of the hot gas must be ejected out of the halo reducing the hot gas reservoir available for gas cooling \citep{guo2011, henriques2013, hirschmann2016, cattaneo2020}. Hence, SAG also considers the energy conservation argument presented by \citet{guo2011} to calculate the ejected hot gas mass which SAG models in a way similar to the modified reheated mass, as also done by \citet{hirschmann2016}.
                    \item \FracReinc:
                    In the ejection scheme used here, the cold gas reheated by SNe explosions is expelled from the galactic disc and stored in an external reservoir. This material that leaves the halo is reincorporated into the hot halo gas on a timescale which depends on the virial velocity of the host halo. SAG introduces the factor that involves the virial velocity following \citet{guo2011}, thus taking into account the fact that the mass ejected by lower mass systems is likely more difficult to be re-accreted since the wind velocities are higher relative to the escape velocity.
                    \item \PertDist:
                    A galactic disc that becomes unstable and is also perturbed by a neighbouring galaxy will undergo a starburst; stars created in this event contribute to the bulge formation. Matter from the stellar disc is also transferred to the bulge. SAG assumes a galaxy to suffer the effects of the interaction when the mean distance between galaxies sharing the same DM halo is smaller than \PertDist~ times the disc scale length of the unstable galaxy. The effect of disc instability in the calibration process is regulated by \PertDist.
                    \item \FracBH:
                    A SMBH grows via gas flows to the galactic centre triggered by the perturbations to the gaseous disc which result from galaxy mergers or disc instabilities. When a merger occurs, central SMBHs are assumed to merge instantaneously. In the case of disc instabilities, only the host galaxy is involved, regulating the fraction of cold gas accreted onto the central SMBH.
                    \item \KAGN:
                    A cooling flow occurs once a static hot gas halo has formed around the central galaxy, and is assumed to be continuous. In our calibration \KAGN~ regulate efficiency of cold gas accretion onto the SMBH during gas cooling.
                    \item \AlphaRP:
                    We consider the model for ram pressure stripping of cold gas disc introduced in \citet{tecce2010}, which is based on the simple criterion proposed by \citet{gunn1972}. The cold gas of the galactic disc located at a galactocentric radius $R$ will be stripped away if the RP exerted by the ambient medium on the galaxy exceeds the restoring force per unit area due to the gravity of the disc.
                    \item \EtaSN:
                    This parameter represents the number of supernovae per unit of stellar mass formed. In SAG, this value directly influences the amount of energy injected by SNe, affecting the heating and ejection of cold gas in the galaxy.
                \end{itemize}

\begin{figure}
                    \centering
                    \includegraphics[scale=0.44]{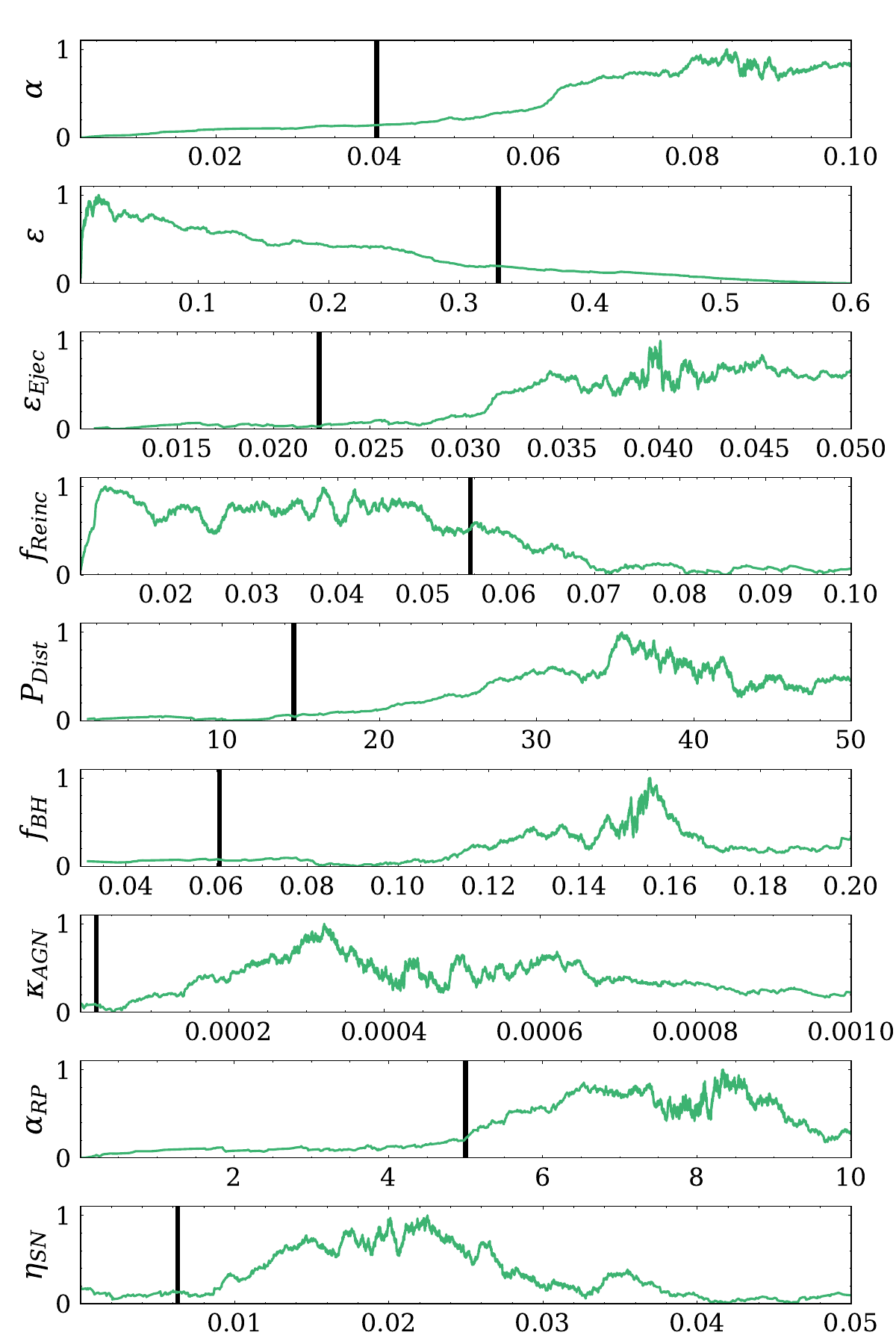}
                    \caption{Likelihood for each parameter in the 7K-SAG calibration. The x-axis range shown in each panel corresponds to the range used in the calibration for each parameter. The vertical black line corresponds to the value obtained in the SAG calibration using the subset of DMO MultiDark simulation \citep{klypin2016}.}
                    \label{fig: sam parameters sag: standard and calibrated}
                \end{figure}

                In Fig. \ref{fig: sam parameters sag: standard and calibrated}, we present the likelihood distributions for each parameter obtained during the 7K-SAG calibration process, representing the probability density that the calibrated value is the optimal one (with a likelihood value of 1). The green lines represent the likelihoods of each optimized parameter, while the black vertical lines mark the standard values of each parameter, as derived from a subset of the DMO MultiDark simulation \citep{klypin2016}. The ranges over which the free parameters were allowed to vary during calibration are also shown. The likelihood distributions exhibit well-defined peaks for most parameters, indicating a stable optimisation process. Parameters like \Alpha~ and \EpsilonEjec~ show sharp peaks, highlighting their key roles in star formation and gas ejection, while others, such as \PertDist, display broader peaks, suggesting a wider range of viable values that still align well with 7K-GIZMO.
                
    \end{appendix}

\end{document}